# The *FarView* Low Frequency Radio Array on the Moon's Far Side: Science and Array Architecture

Jack O. Burns[1], Judd Bowman[2], Tzu-Ching Chang[3], Gregg Hallinan[5], Alex Hegedus[4], Nivedita Mahesh[5,1], Bang Nhan[6], Jonathan Pober[7], Ronald Polidan[8], Willow Smith[7], Nithyanandan Thyagarajan[9]

[1]Center for Astrophysics and Space Astronomy, University of Colorado Boulder, 593 UCB, Boulder, CO 80309
[2]School of Earth and Space Exploration, Arizona State University, ISTB4-675, Mail Code 6004, Tempe, AZ 85287
[3]Jet Propulsion Laboratory, California Institute of Technology, MS 138-308, 4800 Oak Grove Drive, Pasadena, CA 91109
[4]Space Science Institute, 4765 Walnut St., Suite B, Boulder, CO 80301
[5]Cahill Center for Astronomy and Astrophysics, California Institute of Technology, MS 249-17, Pasadena, CA 91125
[6]National Radio Astronomy Observatory, 520 Edgemont Road, Charlottesville, VA 22903
[7]Department of Physics, Brown University, Box 1843, Providence, RI 02912
[8]Lunar Resources, Inc., 18108 Point Lookout Drive, Houston, TX 77058
[9]CSIRO, Space & Astronomy, P. O. Box 1130, Bentley, WA 6102, Australia

## 1.0 Executive Summary

*FarView* is a transformative low-frequency radio observatory proposed for deployment on the far side of the Moon, uniquely enabled by the lunar radio-quiet environment and emerging in-situ resource utilization (ISRU) capabilities. Operating over the ∼1–50 MHz band inaccessible from Earth, *FarView* will open the last unexplored electromagnetic window for astrophysics, delivering breakthrough science across cosmology, heliophysics, Galactic astronomy, and exoplanet habitability.

The primary science driver for *FarView* is the exploration of the Cosmic Dark Ages, the pristine epoch between recombination and the formation of the first stars. Identified by the **Astro2020 Decadal Survey** as the *Discovery Area for Cosmology*, this era is inaccessible to all existing and planned ground- and space-based observatories. By measuring the redshifted 21-cm neutral hydrogen signal from redshifts $z \approx 30–100$, *FarView* will provide three-dimensional tomographic maps and precision power spectra of the early Universe in a largely linear regime. These observations will enable unprecedented tests of fundamental physics, including inflationary initial conditions, primordial non-Gaussianity, the nature of dark matter (including ultra-light axions and dark matter–baryon interactions), neutrino masses, and early dark energy.

*FarView's* reference architecture consists of ~100,000 crossed-dipole antennas, grouped hierarchically into beamformed subarrays and subarray clusters, distributed in a dense core–halo configuration spanning approximately 200 km². The compact core (~4 km across, ~83,000 dipoles) maximizes sensitivity to large-scale cosmological modes, while ~20,000 halo antennas extending to ~14 km provide the angular resolution and calibration fidelity required for foreground characterization and subtraction. Sensitivity analyses demonstrate that this architecture achieves a $\gtrsim 10\sigma$ detection of the Dark Ages 21-cm power spectrum at $z \approx 30$ over a five-year observing campaign, assuming half-duty-cycle lunar-night operations. A trade study of signal processing approaches identifies an FFT-based *EPIC* beamformer as the optimal solution, minimizing computational and data-volume demands while preserving scientific performance.

Beyond cosmology, *FarView* is a powerful multi-disciplinary observatory. Its mixture of baselines and high sensitivity enable direct interferometric imaging of low-frequency solar radio bursts, addressing key priorities of the **Heliophysics Decadal Survey** by constraining particle acceleration, shock formation, and turbulence in the inner heliosphere. These observations will significantly advance space-weather forecasting, a critical element of NASA's **Moon-to-Mars objectives**. *FarView* will also probe stellar space weather in nearby systems, detecting coherent radio bursts from other stars and providing constraints on stellar coronal mass ejections beyond the Sun.

In Galactic astrophysics, *FarView* will perform cosmic-ray tomography of the Milky Way through ultra-long-wavelength measurements of free-free absorption, yielding the first three-dimensional maps of energetic electrons and magnetic structure in the interstellar medium. At the frontier of exoplanet science, *FarView's* sensitivity below 50 MHz enables the detection of auroral radio emission from exoplanet magnetospheres, a key diagnostic of atmospheric retention and habitability, particularly for planets orbiting active M-dwarf stars.

With rapid expansion of human, robotic, and commercial activity at the Moon, the lunar far side's unparalleled radio quietness is a finite and vulnerable resource. Early deployment of *FarView* pathfinders and phased construction of the full array are therefore time-critical. *FarView* represents a flagship-class opportunity to establish the Moon as a platform for foundational astrophysics while delivering discovery-level science that cannot be achieved anywhere else in the foreseeable future.

## 2.0 The Dark Ages – The Discovery Area for Cosmology

*The neutral hydrogen 21-cm line provides a unique probe of the Universe's Dark Ages ($z \gtrsim 30$), a crucial yet unexplored era between recombination that released the CMB photons and the ignition of the first stars. It lies beyond the reach of JWST, Roman, HERA, and other current facilities. Observations at 5–50 MHz require a radio-quiet platform above Earth's ionosphere, motivating lunar far-side experiments: LuSEE-Night will deliver the first global constraints from the Moon, while a future interferometer such as FarView could map 3D matter fluctuations across $z \sim 30–100$ in a largely linear, astrophysically pristine regime. In this section, we describe how such surveys would access orders of magnitude more modes than the CMB or galaxy surveys, enabling ultra-precise tests of inflation (via primordial non-Gaussianity), powerful discrimination between cold and fuzzy dark matter, sensitivity to dark matter–baryon scattering, and high-precision reconstruction of the matter power spectrum. Together, these measurements would provide an unparalleled laboratory for testing ΛCDM, probing the inflationary paradigm and the nature of dark matter, and informing emerging hints of time-evolving dark energy.*

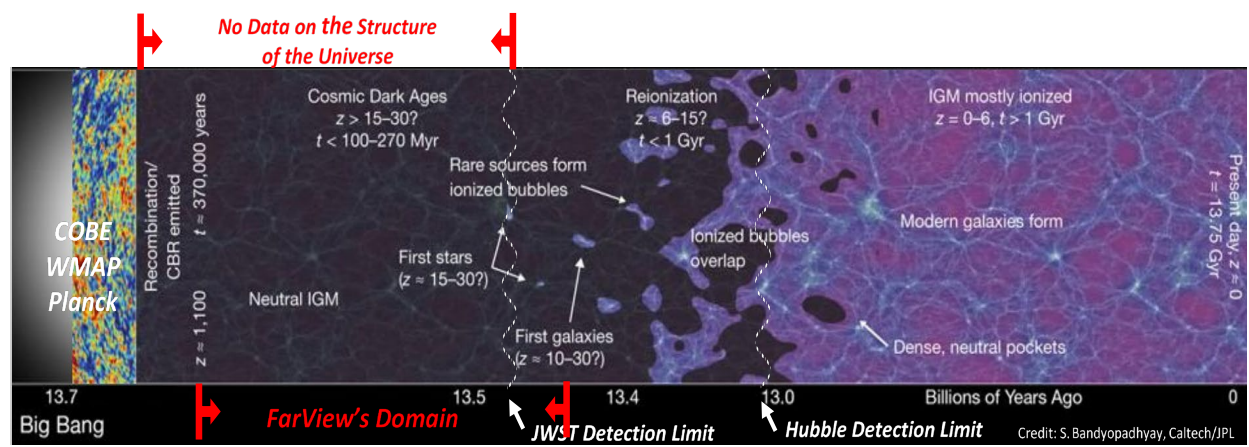

**Figure 1.** The pre-stellar (Dark Ages), first stars (Cosmic Dawn), and Reionization epochs of the Universe can be uniquely probed using the redshifted 21-cm signal. This history is accessible via the neutral hydrogen spin-flip background. *FarView* will fill in the missing data during the Cosmic Dark Ages and Cosmic Dawn.

## 2.1 Overview of 21-cm Cosmology in the Dark Ages

There is a profound gap in our understanding of the early Universe between the epochs of Recombination and Reionization. Figure 1 illustrates this knowledge gap. After the Big Bang, the Universe was hot, dense, and nearly homogeneous. As the Universe expanded, the material cooled, condensing after ~400,000 years (redshift z~1100) into neutral atoms (Recombination), freeing the Cosmic Microwave Background (CMB) photons. There were no compact sources of radiation during this pre-stellar "Dark Ages". The baryonic (normal matter) content consisted primarily of neutral hydrogen (HI) distributed throughout the intergalactic medium. Some 50-100 million years later, gravity initiated the formation of the first luminous objects – stars, black holes, and galaxies – which ended the Dark Ages and commenced the "Cosmic Dawn" [1]. These mostly metal-free first stars (Population III or Pop III) likely differed dramatically from stars we see nearby (Pop I and II), as they formed in vastly different environments [2].

This transformative event marked the first emergence of structural complexity in our Universe, but no currently planned mission can make observations this far back in time. The first data released from *JWST* point toward potentially important new insights into early galaxy formation [e.g., 3, 4, 5, 6]. *JWST*, the *Roman Space Telescope*, and a suite of ground-based facilities will observe the Universe as it was $\gtrsim$ 300 Myrs after the Big Bang (z $\lesssim$ 15, and therefore especially focus on the Reionization era when distant galaxies ionized the gas between them, up to about a billion years after the Big Bang). However, none now contemplate observing the true first stars and black holes [7]—much less the Dark Ages that preceded them. For example, CMB observations of Thomson scattering measure the integrated column density of ionized hydrogen, but only roughly constrain the evolution of the intergalactic medium [8]. Ly$\alpha$ absorption from quasars only confines the end of reionization at relatively late times, z~7 or 770 Myrs after the Big Bang [9]. Observations with the *Hubble Space Telescope* and *JWST* will simply find the brightest galaxies at high redshifts (z$\lesssim$15), and thus any inferences drawn about the high-z galaxy population depend upon highly uncertain assumptions about the faint-end slope of the luminosity function [10]. The *Hydrogen Epoch of Reionization (EoR) Array* (HERA; [11]) is attempting to observe the neutral hydrogen angular power spectrum over a bandwidth of ~50–200 MHz (z = 27 − 6, 117 − 942 Myrs) covering the EoR but not the Dark Ages [12] which is not accessible from the ground. *SPHEREx* [23] will study the last phases of Cosmic Dawn and the EoR (z~6-10) using intensity mapping. Together, all the above observations will provide complementary probes to the 21-cm spin-flip signal and will help to optimize the design of lunar radio telescopes [13].

The 21-cm line of neutral hydrogen offers the only known observational probe of the Universe's Dark Ages. This technique requires observations at wavelengths longer than that allowed by Earth's ionosphere and must be free of anthropogenic radio frequency interference, necessitating a mission on the Moon's far side. Neutral hydrogen can be detected through its redshifted hyperfine 21-cm transition: this signal from the cosmic Dark Ages (redshifts z$\gtrsim$30) is observed at decameter wavelengths or frequencies ~5-50 MHz.

The first observations corresponding to the frequency band for the Dark Ages (1-50 MHz) will be conducted by *LuSEE-Night* from a NASA commercial lander on the lunar far side beginning in Q1 2027 [42, 43]. With batteries allowing night-time observations, we expect the mission to last up to two years. With a single cross-dipole antenna, spectral observations will provide the first constraints on the all-sky 21-cm global monopole from the Dark Ages (Figure 4).

By measuring the signature over a wide range of frequencies **and** angular scales with a radio interferometer, one enables a three-dimensional reconstruction of the distribution of matter (since redshift can be mapped to distance). Thus, with *FarView*, the ultimate deliverable of HI cosmology can be realized - a tomographic map tracing the evolution of the Universe across a huge swath of cosmic history stretching from before the birth of the first luminous objects through the early Cosmic Dawn (first stars) epoch. It is worth reiterating that ***the Dark Ages and the earliest portion of Cosmic Dawn are only accessible from the Moon's far side***.

Figure 2 illustrates the expected evolution of structures in the early universe for the standard ΛCDM cosmology model as viewed by the redshifted 21-cm signal - the amplitude of spatial fluctuations for a range of structure modes. The simplest way to quantify these fluctuations is with the power spectrum, which characterizes the amplitude of the variations as a function of spatial scale, analogous to *Planck* measurements of the CMB. During this time, the 21-cm line traces the cosmic density field with most eigenmodes in the linear or mildly non-linear regime, allowing a straightforward interpretation of the measurement in terms of the fundamental parameters of our Universe [14, 15]. The lack of luminous astrophysical sources makes the Dark Ages signal a clean and powerful cosmological probe and renders the 21-cm line the ***only*** observable signal from this era. Any departure from these well-constrained predictions in Figure 2 will provide important new insights into potentially new and possibly exotic physics during the Dark Ages.

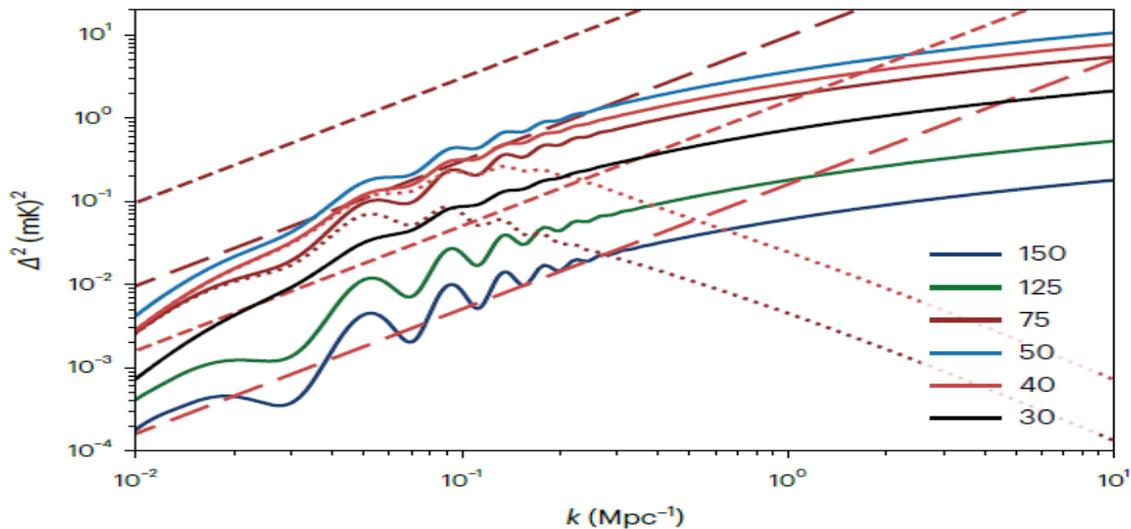

**Figure 2.** The mean squared amplitude for each scale of structure in the early universe forms the power-spectrum as a function of structure wavenumber ($k$). The power spectrum is calculated from the Fourier transform of the matter overdensity field relative to the average density. The colors show the power spectrum at different redshifts for the ΛCDM cosmological model. The "wiggles" at high redshift are caused by acoustic oscillations (i.e., sound waves). The power spectrum is one of the major measurements to be made by *FarView*. Figure is from [16].

## 2.2 Ultimate Tests of Cosmology and Fundamental Physics with *FarView*

The CMB has thus far provided the most stringent constraints on our cosmology model from a 2-D thin shell at z~1100, limited by Silk damping [17] at small angular scales which blurs out fluctuations in the CMB. On the other hand, for the Dark Ages 21-cm power spectrum, the number of accessible modes is enormous and is effectively only limited by the collecting area of the instrument. Late-time galaxy surveys trace the 3-D large-scale structure but are constrained by nonlinear evolution and bias at low redshift; in contrast, the 21-cm signal from the Dark Ages provides a 3-D map of linear density fluctuations across redshift z~30–100. It offers access to a pristine, linear regime across a wide range of scales, largely free of complications from baryonic astrophysics, and is ideal for probing fundamental cosmology. In principle, full-sky 21-cm surveys with arcminute resolution and kilohertz spectral bandwidth could access up to $10^{12}$ independent modes, compared to $10^7$ in the CMB and $10^8$ in current galaxy surveys [18]. This enormous mode count enables exquisite statistical power to test fundamental physics, including the nature of dark matter, the small-scale matter power spectrum, and the structure of inflationary perturbations, with unprecedented precision.

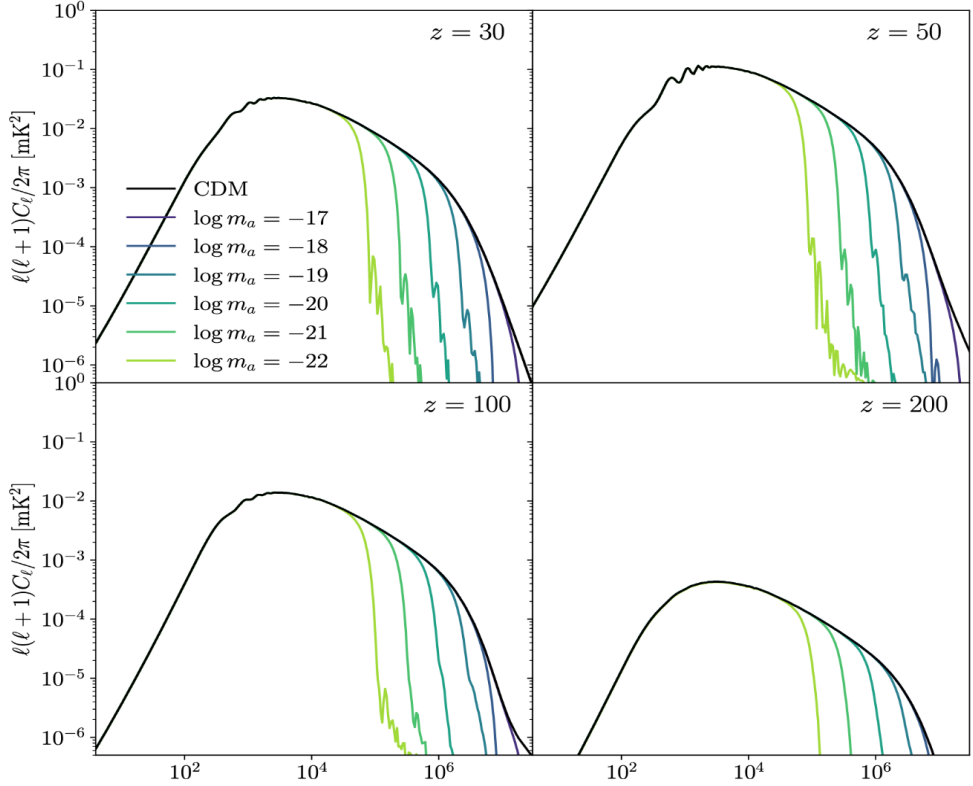

**Figure 3.** The expected 21-cm angular fluctuation power spectrum during Dark Ages, demonstrating the dependance on axion mass ranging from 10-17 to 10-22 eV (color curves), on the small-scale power spectrum at redshift of (30, 50, 100, 200) compared to CDM (black curve; [29]). Without the complication of baryonic physics, the interpretation of the measured small-scale 21-cm fluctuation power would place tight constraints on the nature of dark matter. $l$ is the spherical harmonic moment (where large $l$ corresponds to small angular scales). This figure is from [29].

Fluctuation measurements of 21-cm brightness temperature provide a powerful probe of inflationary physics. Inflation, the leading paradigm for the universe's rapid early expansion, accounts for its observed flatness, homogeneity, and isotropy. During this epoch, quantum fluctuations were stretched to cosmological scales, seeding the density perturbations that later evolved into large-scale structure. A key prediction is that these fluctuations are nearly Gaussian and scale-invariant. However, small deviations from Gaussianity, parametrized by $f_{NL}$, can reveal the number, dynamics, and interactions of the fields driving inflation [19]. Precise measurements of $f_{NL}$, thus offer a direct probe of the inflationary process, enabling sharp discrimination between single-field and multifield models, and shedding light on the fundamental physics that governed the earliest moments of the universe.

The enormous statistical power of the Dark Ages 21-cm fluctuations enables sensitive measurements of $f_{NL}$, with forecasts suggesting sensitivities of $\Delta f_{NL} \sim 0.01–0.03$ under realistic conditions, and down to 0.006 nearing the cosmic variance limit with cross-correlation with the CMB [16, 20, 21]. These are orders of magnitude tighter constraints than current generation expected by the Dark Energy Spectroscopic Instrument (DESI) [22] galaxy survey or the NASA *SPHEREx* mission [23], which offer $\Delta f_{NL} \sim 1$. ***The Dark Ages thus has the potential to provide the ultimate constraint on inflationary sciences.***

In addition, the nature of dark matter, which makes up roughly 25% of the present energy budget of the Universe, is one of the key unanswered questions in cosmology. While the standard cold dark matter (CDM) scenario has been well tested on large scales [24], there remain inconsistencies below galactic scale observations [25] hinting a lack of power on small scales with respect to the standard ΛCDM cosmological model. An appealing alternative candidate is the ultra-light axion dark matter (ULAs) or fuzzy dark matter (FDM), which consists of dark matter being made of ultra-light scalar fields with masses in the range $10^{-27} \sim 10^{-10}$ eV [26, 27, 28]. This scenario provides the same large-scale predictions as CDM, but the wave-like nature of the fields naturally suppresses small-scale structure. Figure 3 shows the corresponding 21-cm 2-D angular power spectrum as a function of redshift, for different values of the axion mass compared to the CDM scenario [29]. Clearly, the axion mass-dependent suppression of small-scale (large spherical harmonic moment *l*) power is evident and crucially, it can be cleanly discerned and interpreted at high-redshift without the complication of baryonic physics that also impacts small-scale structure.

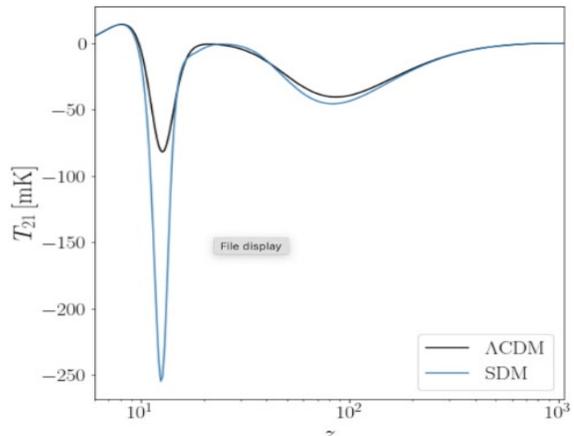

**Figure 4.** The globally averaged mean 21-cm brightness temperature as a function of redshift, shown in standard ΛCDM model (black curve) and a DM-baryon scattering dark matter model (SDM). From [30].

Other exotic DM models include the DM-baryon scattering model, where dark matter particles elastically scatter off standard model particles, resulting in heat flows from the hotter baryons to the colder DM particles. Additionally, the friction between the fluids results in a drag force that tends to lower the relative velocity as well as heating up both fluids. This results in an enhanced absorption feature of the 21-cm mean temperature against the CMB during Cosmic Dawn and the Dark Ages. There has been an enhanced interest in DM-baryon scattering model to interpret the EDGES result, reporting a deeper than expected absorption trough at z~17 during Cosmic Dawn [32]. As demonstrated in Figure 4 [30], this makes 21-cm signals uniquely sensitive to the microphysics of DM particles at high redshift. Moreover, the effect of DM-baryon scattering on the 21-cm fluctuation power spectrum is also evident, as shown in Figure 5 [31] as a function of redshift for different DM mass. This again demonstrates the unique constraining power of high redshift 21-cm fluctuation measurements on the nature of dark matter.

Furthermore, the fine scale of 21-cm measurements allows reconstruction of the full matter power spectrum across a wide range of scales, offering complementary information to large-scale structure and CMB to measure and test our cosmological model [15]. This is of particular interest given the recent DESI results hinting on a non-constant, time-evolving dark energy [33]. While dark energy domination is a late-time phenomenon, it will nonetheless be imperative to map out the energy content of the Universe across a wide range of redshifts to firmly test and understand the fundamental physics and the cosmological model.

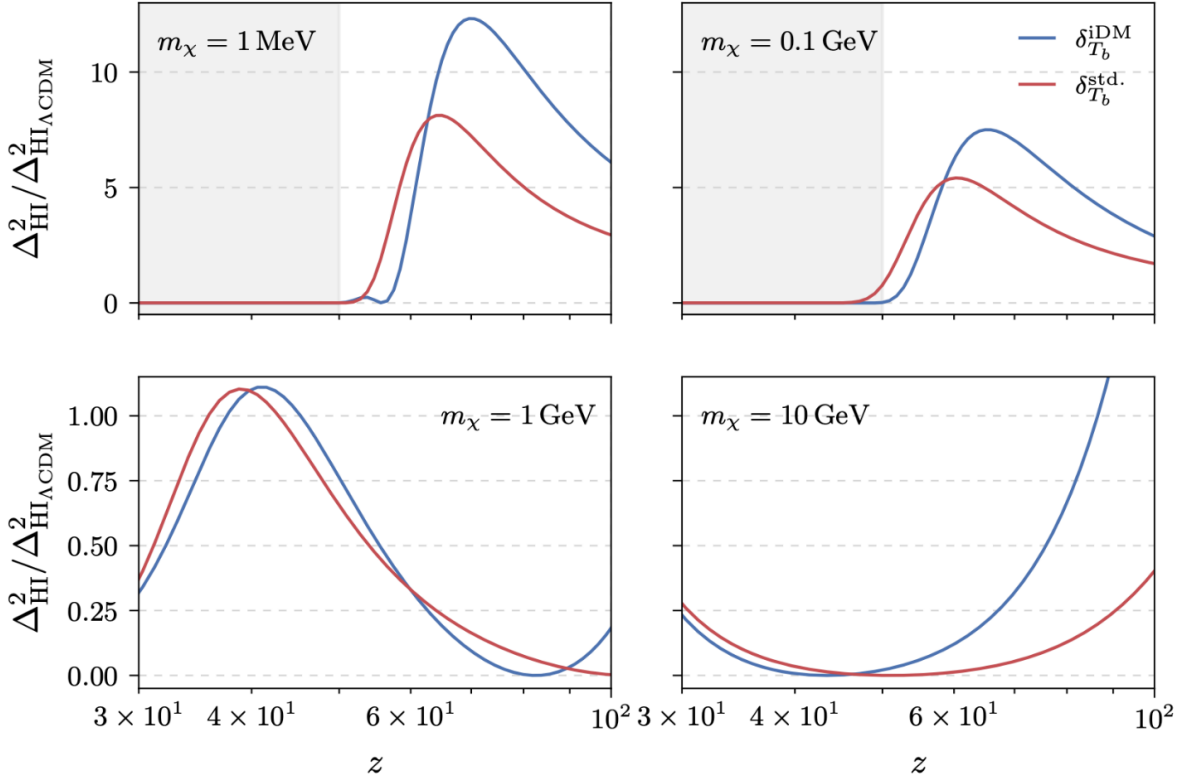

**Figure 5.** The ratio of the 21-cm power spectra at k= 1 Mpc$^{-1}$ with respect to ΛCDM from [31]. Blue curves properly include the additional effect of temperature fluctuations due to DM-baryon scattering while red curves ignore it. This shows that at DM mass less than 1 GeV, there is a large enhancement at high redshifts, peaking at z ~70; although in these cases, the cooling of the baryon gas temperature drives the mean brightness temperature towards zero at z < 50. For a 1 GeV DM mass, the amplitude of the 21-cm power spectrum is enhanced for z > 40 due to the enhanced baryon temperature fluctuations, except close to z ~80 due to the suppression of the baryon temperature fluctuations. For a 10 GeV DM the 21-cm power spectrum is enhanced z > 50.

### 2.3 Community Recommendations for Dark Ages Cosmology from the Moon

*FarView* addresses ***NASA's objectives in astrophysics***[1] to "discover how the universe works, explore how it began and evolved" and to "probe the origin and destiny of our universe, including the nature of black holes, dark energy, dark matter and gravity." It will execute a mission recommended in the ***NASA Astrophysics Roadmap*** [68] to observe "the Universe's hydrogen clouds using 21-cm wavelengths via observations from the farside of the Moon." The ***NRC Astro2020 Decadal Survey*** [44] singled out the Dark Ages as "the Discovery Area in cosmology"; and identified "needed capabilities to include next-generation 21-cm interferometers… with both higher sensitivity and a better understanding of instrumental systematics and astrophysical couplings." The *Artemis III Science Definition Report*[2] advocated for observations from "the farside of the Moon and the 21-cm electromagnetic radiation spectral line to study the Dark Ages of the universe." Finally, *NASA's Moon to Mars objectives*[3] include conducting radio observations from the Moon's far side to study the cosmic Dark Ages using radio quietness.

---
[1] https://science.nasa.gov/astrophysics/

## 3.0 The Architecture and Sensitivity of FarView for Dark Ages Observations

*The precise architecture of the FarView array is paramount to observing the cosmic Dark Ages which by its faint nature requires a very large dipole antenna array. Herein, we present an evolution on the FarView architecture first described in [58] with what we call the fiducial FarView array. In short, the fiducial FarView array has a densely packed core of 80,000 10-m crossed dipole antennas (along with 20,000 outrigger antennas) that offers a $10\sigma$ detection of the redshift z=30 power spectrum over a 5-year half-duty cycle – i.e. the array does not operate when the radio bright sun is up – in the absence of foregrounds. Dipoles are grouped into subarrays of 16 dipoles each which are beamformed together into one receiving element, and subarrays are further grouped into clusters of 36 subarrays which connect to a solar battery. We also vary parameters determining spacing and layout of FarView subarrays, where we find and emphasize that a densely packed array is necessary to achieve any detection of the Dark Ages power spectrum.*

### 3.1 The Fiducial Array

*FarView* is an in-situ manufactured radio interferometer capable of a number of scientific investigations (see Sections 2 and 5). In this section, we examine *FarView's* primary goal of observing the cosmic Dark Ages power spectrum and therefore focus on what we call the fiducial array. Importantly, *FarView* is designed to be constructed iteratively and can begin some of its science goals before reaching the stage described as the fiducial array here.

The initial *FarView* concept and layout was first presented in [58] published in *Advances in Space Research* (ASR) where it will henceforth be referred to as the ASR concept. The ASR concept introduced a number of key aspects and specifications to the architecture of *FarView*. Namely, the ASR concept presented a ~100,000 element array with half the elements being in the core array with the remaining half being outriggers. Each element, a single dipole with ≈30 m² collecting area, is grouped into a 500-element *subarray* which, through the process of beamforming, acts effectively as a single antenna. The implementation of *subarrays* is to address the computationally demanding, if not impossible, task of correlating pairs of dipoles which for an array of the size necessary to observe the Dark Ages is on the order of $N^2 \sim 10^{10}$ correlations where N is the number of dipoles (see Section 8).

In [59], it was found that the ASR concept lacks sufficient sensitivity to detect the z=30 21-cm power spectrum over a 5-year observation at half-duty cycle. The following changes were therefore made to the *FarView* architecture: all dipoles are now crossed dipoles, subarrays are now 16 dipoles in total and spaced 11.5-m apart, and 36 subarrays now comprise a single *cluster* – explicitly defined by the number of subarrays that are supplied power by a single solar battery. Additionally, the dipole split between core and outriggers now favors the core with 82,944 dipoles or 144 *clusters* in the core and the remaining clusters as outriggers. For sensitivity analysis presented later, outriggers – which are useful for calibration but contribute little to the detection of the cosmological signal – are ignored. A schematic diagram of the core, henceforth referred to as the fiducial array, is presented in Figure 6 and achieves a 10$\sigma$ detection of the z=30 21-cm power spectrum over a 5-year observation at half-duty cycle in the absence of foregrounds.

---

[2] https://www.nasa.gov/wp-content/uploads/2015/01/artemis-iii-science-definition-report-12042020c.pdf?emrc=841cb1
[3] https://www.nasa.gov/wp-content/uploads/2022/09/m2m-objectives-exec-summary.pdf

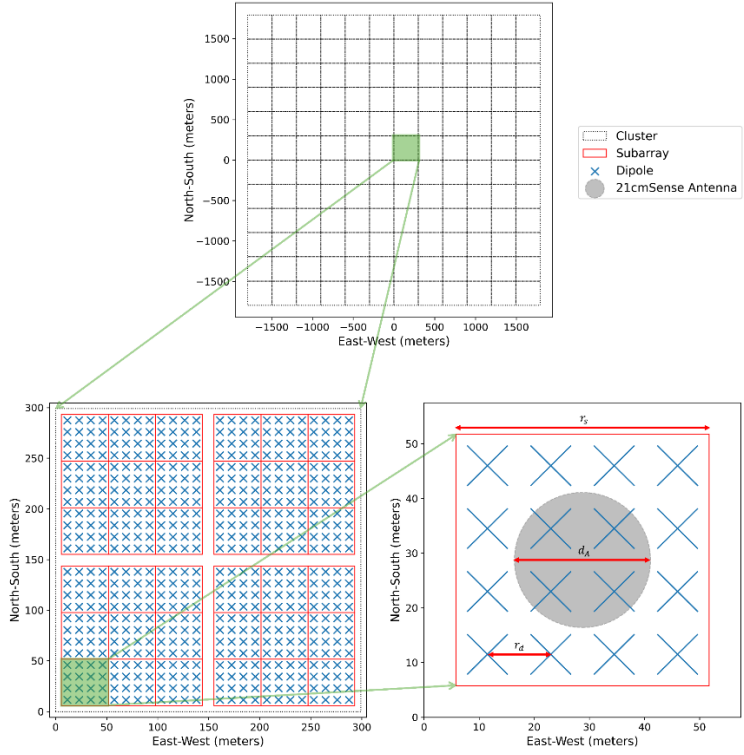

**Figure 6:** The fiducial array core. *Top:* Each square represents a cluster of 36 subarrays. *Bottom Left:* Zoom-in on a cluster where each red square is a subarray. *Bottom Right:* Zoom-in on a subarray which consists of 16 crossed 10-m dipoles; the grey circle is the equivalent single aperture antenna ($d_A$=25-m diameter) simulated by *21-cmSense*.

## 3.2 Calculating Sensitivity

In this section, we briefly describe how the sensitivity analysis of *FarView* is performed (for more detail see [59]).

### 3.2.1 The 21-cm Power Spectrum

To calculate the sensitivity of *FarView*, we first generate a brightness temperature power spectrum at z=30. We do this in *21-cmFast* [60] using a 2 Gpc box with a 4 Mpc voxel size, spin temperature fluctuations enabled, and the default parameters of *21-cmFast* v3.3.1. This brightness temperature power spectrum when normalized differs from the normalized dark matter density power spectrum by 3.5% – indicating that ionization, heating, and velocities play a subdominant role as drivers of the brightness temperature. Thus, we use the brightness temperature power spectrum for all sensitivity calculations while keeping in mind that it is a direct parallel to the dark matter density.

### 3.2.1 The 21-cm Noise Power Spectrum

The noise contribution to the 21-cm power spectrum for a radio interferometric array was first derived in rigorous detail in [61] and [62] and is often referred to as the 21-cm noise power spectrum. The contributions to the noise power spectrum can be described in two types: thermal variance (characterized by the radiometer equation) and sample variance (typically referred to as cosmic variance). However, at Dark Ages frequencies, the sky brightness temperature is substantial, making the thermal variance several orders of magnitude larger than the sample variance – in the case of the fiducial array, the thermal variance at best is 4 orders of magnitude larger than the sample variance. Therefore, we ignore the subdominant sample variance in our sensitivity analysis for *FarView*.

For white-noise thermal fluctuations with root mean square (rms) brightness temperature, as shown in [63], the thermal noise contribution to the dimensionless power spectrum $\Delta^2(k)$ can be written as

$$\delta\Delta^2(k) \approx X^2 Y \frac{k^3}{2\pi^2} \Omega T_{rms}^2$$

Here $X^2Y$ is the cosmological scaling factor between observing coordinates or comoving volume, $\Omega$ is the primary beam field of view – as described in Appendix B of [64] – B is the observing bandwidth, $k$ is the Fourier mode of the one-dimensional 21-cm power spectrum, and $T_{rms}$ is the effective radiometric noise temperature for one $k$ mode. Here we emphasize that $T_{rms}$ is proportional to variations in the architecture of *FarView* while all other factors remain constant.

Using the package *21-cmSense* [65], we can calculate the noise spectrum for a lunar array like *FarView*. We make one simplification, treating each subarray as a single circular dish antenna with equivalent effective area $A_{eff}$ to a subarray as seen in Figure 6. We justify this simplification by the fact that $T_{rms}$ is only concerned with total area and not the precise shape of the antenna.

We show the noise power spectrum of the fiducial array and z=30 dimensionless power spectrum in Figure 7.

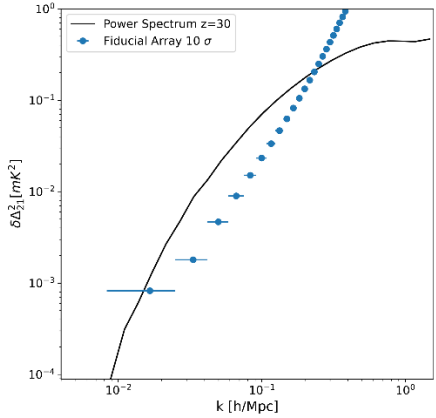

**Figure 7:** One dimensional noise power spectrum (blue) of the fiducial array showing a 10$\sigma$ detection of the z=30 dimensionless power spectrum

### 3.3 Variations of the Fiducial Array

Here we briefly summarize the effects on sensitivity by varying 3 parameters of the fiducial architecture: number of dipoles $n$ in a subarray, the spacing $r_d$ between dipoles, and the total number of dipoles $N$. From these variations we can determine what aspects of array design are most important, namely that increasing the number of small baselines significantly improves the detection.

Here we show 4 variations in $n$ (shown as $\sqrt{n} \times \sqrt{n}$) and 4 variations in $r_d$, where one variation in each is the fiducial array, as shown in Figure 8. In both cases there is an increase in the significance of detection with decreased subarray size i.e. the shortening of the width, $r_s$, of a subarray creates a more densely packed core. Interestingly, $r_s$ of the 6×6 and 3×3 subarrays are roughly equivalent to the $r_s$ of the $r_d = 17$ m and $r_d = 9$ m respectively. However, the 6×6 subarray outperforms its counterpart while the 3×3 subarray underperforms in sensitivity. Understanding that $r_s$ sets the smallest baselines, this shows that for similar baseline distributions, a larger antenna is favored.

We also show the impact on sensitivity from varying the total number of dipoles $N$ or more directly the total number of clusters in Figure 9. This represents *FarView's* ability to expand with the continuous addition of dipoles. We note that each additional cluster represents a 0.08$\sigma$ increase in the detection.

## 3.5 Summary of Sensitivity Analysis

The fiducial array, as described, represents the baseline architecture of *FarView*. With the difficulties presented in constructing a lunar farside array, we believe this represents a good compromise between detection of the cosmic Dark Ages signal and feasibility of design. While it is possible to of course increase sensitivity by increasing observation time or total collecting area, these parameters only trend linearly in sensitivity. The key design parameter, in terms of sensitivity to the dark ages power spectrum, is the number of small baselines which probe large scale structure in the dark matter density.

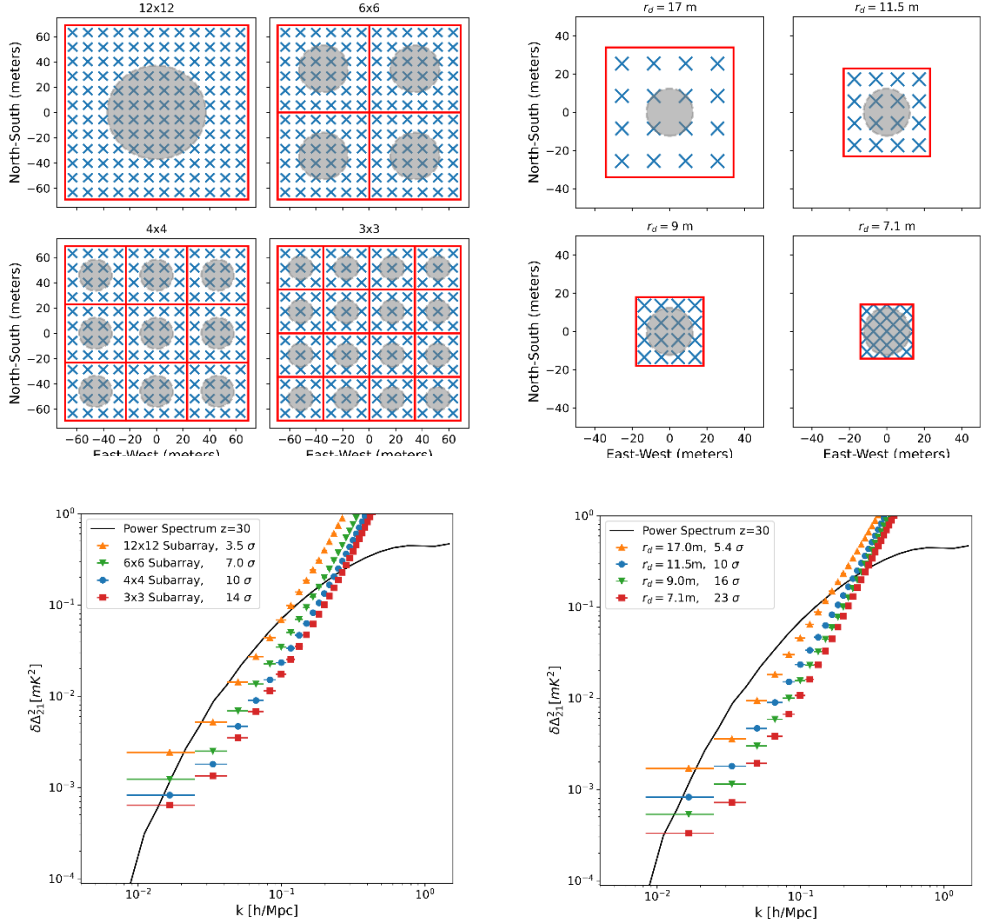

**Figure 8:** Variations on the number of dipoles in a subarray (top left) with respective sensitivity (bottom left) and variations on the dipole spacing (top right) with respective sensitivity (bottom right).

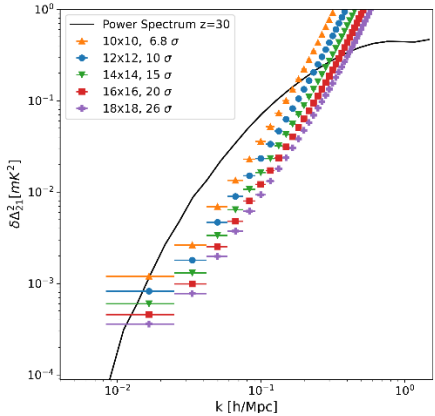

**Figure 9:** The sensitivity offered by variations on the number of clusters in the core written as the dimensions in number of clusters on one side.

# 4.0 Foreground Emission for 21-cm Dark Ages Cosmology

*Bright foreground emission -- and, in particular, the interplay between bright foreground emission and the frequency-dependent systematics of a radio interferometer -- represents the major challenge for 21 cm cosmology. Ground-based experiments targeting a detection of the 21 cm signal from the Epoch of Reionization have developed a framework for isolating these effects ("foreground avoidance") based on the known response of the interferometer. Because Dark Ages experiments target substantially lower radio frequencies, however, the foreground avoidance approach is significantly less effective. Foreground mitigation (likely taking the form of some kind of "foreground subtraction) is therefore an essential area of study and technology development for an experiment like FarView.*

As discussed in Section 6 (on single dish vs. interferometers), a 21-cm cosmology Dark Ages experiment must be built to address the issue of foreground emission. Other astrophysical radio sources — predominantly synchrotron emission from both the Milky Way and extragalactic objects — will swamp the cosmological hydrogen signal by 5–7 orders of magnitude at the frequencies corresponding to the redshifted Dark Ages signal [1]. This foreground emission can, in principle, be separated from the cosmological signal via its distinct spectral behavior (astrophysical foregrounds are spectrally smooth, while the cosmological signal is spectrally structured), but the radio telescope conducting the observations imparts its own spectral signature that complicates the picture.

Much of the approaches developed for foreground mitigation have come from the ground-based community of radio astronomers currently using shorter wavelength observations with interferometric arrays to target the hydrogen signal from the era of first galaxy formation (often referred to as the Epoch of Reionization or EoR). From this community, a popular paradigm of "foreground avoidance" has emerged. In the foreground avoidance approach, a first-principles calculation of the instrument's spectral response leads to a footprint for foregrounds in cosmological Fourier space. Only modes of the cosmological signal falling outside the foreground contaminated region are then used to constrain astrophysics and cosmology. For studies of the EoR, the sensitivity impact is already quite significant, with upwards of 80% of the cosmological modes lost to foregrounds. Nevertheless, given the difficulties associated with a "foreground subtraction" approach (discussed in more detail below), the current best limits on the EoR signal come from using foreground avoidance.

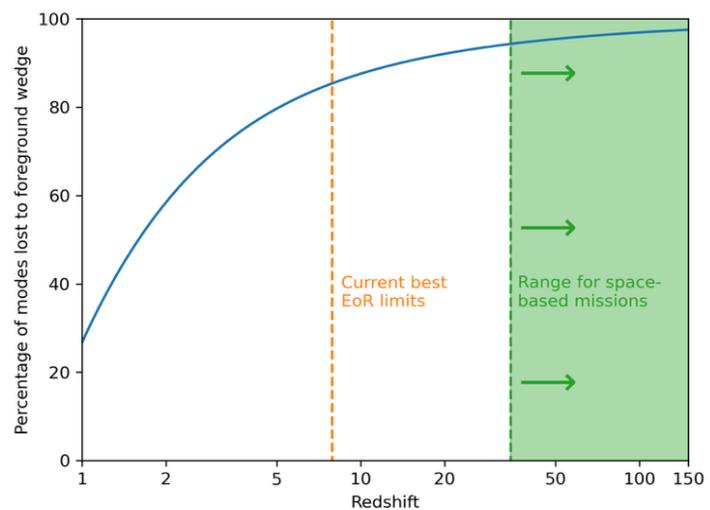

**Figure 10:** The effect of foreground avoidance on the number of useable cosmological modes.

Under-appreciated, however, is that the instrument's footprint in cosmological Fourier space evolves with frequency. At the higher redshifts of the cosmic Dark Ages, an observed angle on the sky maps to a much shorter physical distance than at the EoR; similarly, a fixed bandwidth covers a much larger range of redshifts for the Dark Ages than for the EoR. When these scalings are accounted for, the net effect is that many more modes are lost to foregrounds at the highest redshifts, as illustrated in Figure 10. Given the extreme faintness of the hydrogen signal from the Dark Ages, the effect of foreground avoidance becomes potentially too much to overcome. As discussed in Section 3, an estimated 2.5 square kilometers of core collecting area operating for five years is required to detect the Dark Ages signal. Pober & Smith [34] quantify the impact of the foreground avoidance approach on a Dark Ages experiment and demonstrate that an observatory either 10 times larger or an observing campaign 10 times longer is required to achieve comparable sensitivities to an experiment that can remove foreground contamination. The increasing sky noise towards the higher redshifts for the Dark Ages signal only exacerbates the problem further.

The alternative is, then, to pursue a foreground subtraction approach, where the astrophysical foreground emission is modeled and removed from the data. Such an approach is complicated for ground-based experiments, where the time-variable ionosphere makes accurate foreground modeling extremely challenging. At EoR redshifts, only the LOFAR experiment has pushed foreground subtraction, with an emphasis on developing direction-dependent calibration techniques that can model the ionospheric distortions. On the lunar far side, however, the absence of an ionosphere may greatly simply this requirement; building such a model and recovering many more of the modes of the hydrogen signal may be the most straightforward route to achieving the necessary sensitivity in the near term. A high-fidelity foreground model will require excellent spatial resolution and, in the case of a radio interferometer, *uv*-coverage.

The *FarView* design concept responds to these challenges with its mix of a ~80,000-element dense core (for high sensitivity to the cosmological signal) and ~20,000 outrigger antennas for foreground modeling. While the exact placement for the outriggers is an area for investigation in a potential NIAC Phase 3 study, it is clear from established approaches that the angular resolution of core alone (≈7 arcmin at 45 MHz) is insufficient for a high precision foreground model. Furthermore, the wide field of view enabled by interferometers (set by the individual antenna element size) is a critical component for accurate foreground modeling. Foregrounds well outside the main field of view (and even down to the horizon, see e.g. [35] and [36]) can still swamp the Dark Ages signal, meaning the instrument must be able to produce a high-quality foreground model across the whole sky — a near impossible task for a large, non-steerable single-dish.

## 5.0 Additional Science Cases for FarView

### 5.1 Heliophysics and Space Weather

*This section showcases the advances in Heliophysics that could be conducted by the full Farview array, or even early subsets of it during construction. Solar radio bursts can be the brightest objects in the low frequency sky and accompany extreme space weather like solar flares and coronal mass ejections (CMEs). We use simulated radio emission maps and a simulated FarView array to demonstrate the unprecedented image quality of solar bursts for these low frequencies. These images show where radio emission originates relative to the larger structure of the CME and point to important physics for particle acceleration. FarView would also enable observations of the scattering of these bursts over time, giving us information about the turbulence*

*in the solar wind in the inner heliosphere. It would also progress on **Moon to Mars** goals by advancing our understanding of lunar surface plasma density and its modulations by the Sun and Earth's magnetotail.*

One long-sought goal for low frequency radio observations has been the imaging of low frequency solar radio bursts. There have been many coincident observations of these radio bursts with the space weather events surrounding them, with radio images currently limited to frequencies above the ionospheric cutoff ($\lesssim$20 MHz). It is thought that radio emission marks the site of energetic particle acceleration around the edges of coronal mass ejections and around electron jets from solar flares for Type II and III bursts, respectively [53, 69]. Both Type II and III radio bursts emerge from extreme space weather conditions, where the emitted frequency is related to the local electron plasma frequency, which is driven by the plasma density. Thus, one can take the distance from the Sun and thereby know the emitting fundamental frequency expected at those ranges from heliospheric plasma density models (e.g., [37]).

This research can commence as soon as the first cluster of subarrays is deployed on the lunar surface. Even a single FarView cluster - comprising 576 crossed-dipole antennas - provides sufficient sensitivity and angular resolution to produce high quality interferometric maps of Type II and Type III solar radio bursts. Early observations of these bright, low-frequency solar emissions will play a critical role in validating instrument performance, establishing calibration pipelines, and characterizing the lunar radio environment. As additional clusters are constructed and brought online, the array's instantaneous *uv*-coverage and collecting area will expand, enabling progressively higher-fidelity imaging with improved dynamic range and greater sensitivity to both compact and diffuse solar burst structures.

Many ground-based arrays have tracked these emissions in the low radio frequencies, but due to the Earth's ionosphere, the lower frequency band of 0.1 – 20 MHz is not visible around the Sun during the daytime. Past NASA missions like *STEREO* and *Wind* have made single or multi antenna observations of these emissions, characterizing their spectral and temporal variations. However, these efforts in this frequency range have been limited by the lack of interferometric observations with good spatial resolution. True interferometric data would enable imaging of these radio bursts from 2-20 solar radii in the frequency range 0.1 – 20 MHz. This is the primary science goal of NASA's *SunRISE*, an upcoming GEO Earth-orbiting radio interferometer, where satellites in orbit will drift between 1-10 km from each other. This will allow *SunRISE* to make first-of-a-kind measurements in observing the immediate source of radio bursts around space weather events.

One blind spot to this approach however comes from the fact that an orbiting array of *SunRISE's* size lacks short antenna separations needed to measure highly scattered bursts. It is thought that these solar radio bursts undergo scattering shortly after they are emitted. *Parker Solar Probe* observations have shown for the first time inherent polarized structure in Type III bursts within the first few seconds of emitting, before the signal is washed out from the scattering of the waves from the turbulent media between us the observer and the emission site [70]. The detailed behavior of this scattering can reveal important quantities about the solar wind turbulence, as detailed in [71]. Their models show a clear observational gap that only a lunar-based array could fill, as the scattered bursts would quickly go out of the sensitivity range for a spread-out *SunRISE* type array. This is because SunRISE's cross correlated data are far more sensitive to compact sources in the sky, not the broadly scattered brightness patterns after some 10s of seconds of scattering throughout the background solar wind.

Observed variables regarding this scattering can inform us about the turbulent properties within these emission ranges of 2-20 solar radii, a region of the inner heliosphere that is under-sampled compared to the Earth's neighborhood around 200 solar radii. By observing the rate of scattering in size of each burst, the decay time in power, and the position of the burst, one may infer the mean wavenumber levels of turbulence, as well as the anisotropy factor of $k_\parallel/k_\perp$, showing the asymmetry in the turbulence along the magnetic field. *FarView* would be able to provide best in class measurements of these phenomena, yielding better understanding and even near real time alerts of space weather dynamics.

To understand the expected results from the *FarView* array observing these solar radio emissions, we have constructed a synthetic observation pipeline. Starting with the simulated *FarView* array, we constructed a SPICE [72] kernel to align the lunar surface with the sky, including the Sun to track the solar zenith angle from the simulated array over time. In this way the point source function can be tracked over the course of the lunar day, as the phase center of the interferometer aimed at the Sun shifts over time. By using models of the emission for Type II and Type III bursts, we show that *FarView* could capture highly detailed and high cadence images of these emissions, giving an unparalleled view into these dangerous space weather events.

In what follows, we explore the full and partial *FarView* array tasked with observing Type II and III radio bursts. The Type II ground truth comes from a simulated recreation of a CME, and the areas of the shock exceeding some upstream/downstream density ratios are assumed to be the source of the emission. The *FarView* array is then tasked with imaging this via an interferometric snapshot with no time to conduct aperture synthesis. Additional studies are done on the more numerous Type III bursts, where *FarView's* ability to image the highly scattered emission over time is assessed. Another set of studies were conducted to test the image quality possible from a single/few *FarView* cluster(s), as the array under construction will be able to start observing stronger signals like local space weather emissions before the full array is complete.

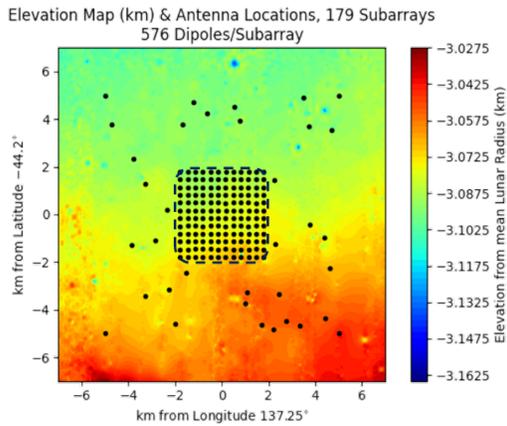

**Figure 11.** 3.6 km core and 14 km diameter halo for *FarView* with an 80-20 split in core-halo antennas (82,944/ 20,160). Each cluster, shown as dots on this figure, shares power and contains 36 subarrays. Each subarray has 16 cross dipoles for a total of 576 cross dipoles/cluster. Background is topographic map from NASA LRO of the Pauli impact basin.

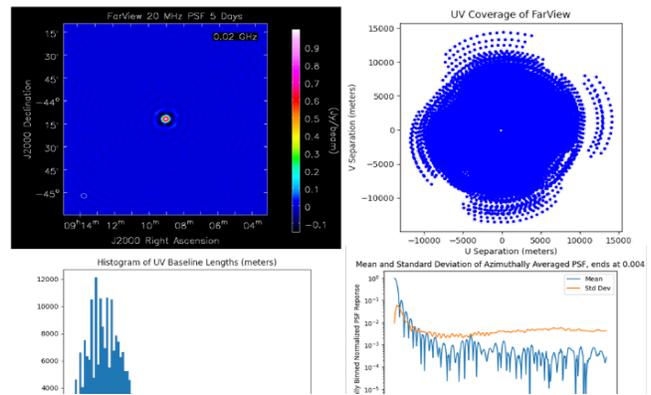

**Figure 12.** *Upper Left:* Point spread function from *FarView's* 179 clusters shown in Figure 11. This simulated observation used an integration over 5 Earth days, 1-5 January 2035, over 15×8 hour discrete steps following the patch of the sky that crosses zenith halfway through. *Upper Right:* uv coverage of this 15 step integration for a phase center at zenith. *Lower Left:* Histogram of baseline lengths for this integration. *Lower Right:* Analysis of point source function in upper left panel, with a sidelobe rms of < 1%.

The core-halo *FarView* design described in Section 3 is illustrated on the Pauli impact basin region on the far side of the Moon in Figure 11; the core has 80% of all the antennas, and packs them tightly together in a 4 km on a side square, and the remaining 20% are in a halo for foreground measurement. This design enables sensitive studies of diffuse 21-cm signals to probe the Dark Ages and has sidelobe levels under 1%, as shown by the analysis plot in Figure 12.

### 5.1.1 Cluster Distribution and Build Sequence

Using the fiducial *FarView* array design, the imaging of these initial arrays was simulated to characterize their expected performance. Figure 13 illustrates the point source function and sidelobe analysis of a single cluster of subarrays. As a reminder, a cluster is composed of 36 subarrays or 576 antennas.

The first few clusters may be built in any order, opening the door for some early studies of brighter solar radio bursts that do not require as many antennas. Figure 14 shows an example of an early 4 cluster configuration situated for maximum distance between clusters. One can see that the point source function of this 4-cluster configuration matches the above pattern for a single cluster, but with a higher resolution pattern superimposed on top.

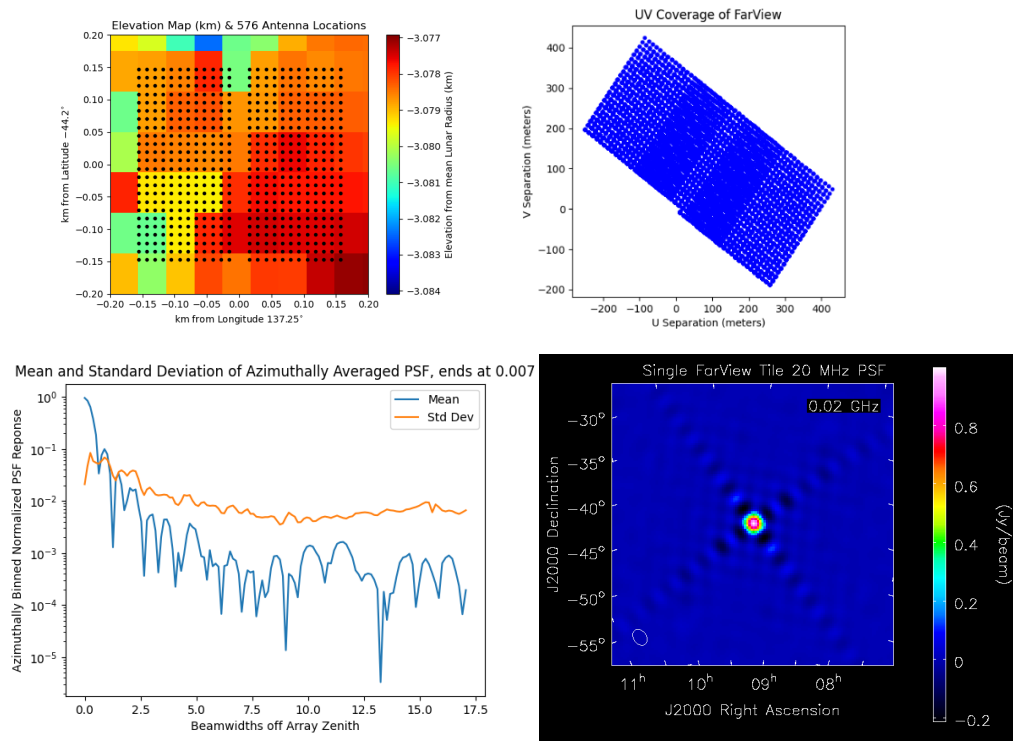

**Figure 13.** *Upper left:* Distribution of antennas for a single cluster over 300 m$^2$. *Upper right:* Instantaneous UV coverage of a single cluster. *Lower left:* Mean and standard deviation of point source function for a single cluster, with x-axis normalized by beamwidth. *Lower right:* Point source function of cluster at 20 MHz from WSClean.

### 5.1.2 Initial Imaging Studies

Using this simulation pipeline described in Hegedus et al. [38], we have made ground truth images of the sky and propagated them through *FarView's* synthetic aperture. After adding the appropriate amount of noise, imaging of the cross correlated visibilities and deconvolution (cleaning) was done with the package wsclean [39]. This demonstrates *FarView's* imaging capabilities across various scientific targets, including 21-cm emission maps and solar radio bursts.

### 5.1.3 Scientific Studies: Solar Radio Burst Imaging

Solar radio bursts are thought to be markers of acceleration process for dangerous solar energetic particles. These bursts occur throughout the inner heliosphere and beyond, with an observational gap under 20 MHz, due to Earth's ionospheric cutoff, and the required size of arrays above the ionosphere to observe and make detailed maps of them. The most common bursts in this frequency range are characterized as Type II or Type III bursts, which are associated with Coronal Mass Ejections (CMEs) and active regions on the solar surface respectively [40]. Basic images of these bursts will be acquired in coming years with NASA's SunRISE mission, which will place six satellites in a grouped GEO orbit, with distances between 1-10 km between receivers [41]. SunRISE's orbit will enable novel measurements, but it will also lack consistently shorter antenna separations. This means that SunRISE is "blind" to larger spatial scale structures and will miss out on highly scattered signals.

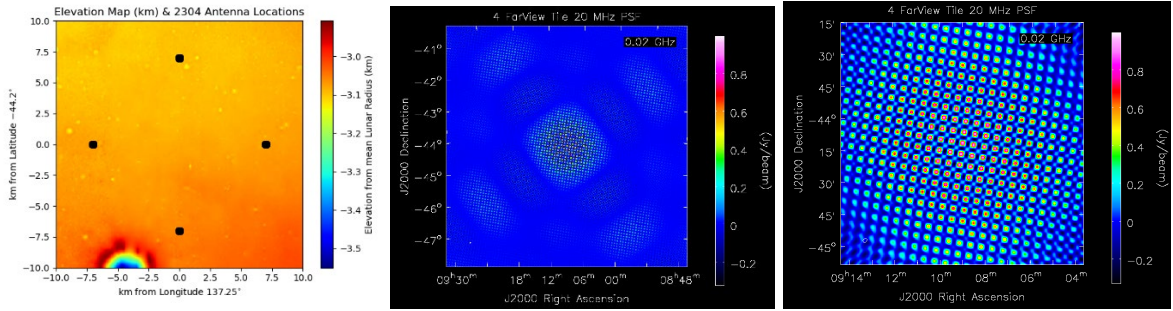

**Figure 14.** *Left:* Distribution of antennas for 4 clusters with full FarView size. *Middle:* Point source function of configuration at 20 MHz from WSClean. *Right:* Zoomed-in point source function to show overlaid high frequency pattern over larger pattern for a single cluster.

The lunar *FarView* array will provide more detailed images of these bursts, using many more antennas (baselines) that together are sensitive to both smaller and larger spatial sizes. This means that FarView could study these solar radio bursts for longer durations, since scattering will lead to radio bursts becoming larger and larger, eventually too large for SunRISE's longer baselines to sense in its cross-correlated data. Figure 15 is an example of a detailed Type II radio burst, along with the recovered image from the simulated *FarView* array. The image is still resolution limited by the wavelength and size of the array but contains much more detail than a six-element interferometer like SunRISE will be able to capture. These high-fidelity images would allow us to compare simulated recreations of the event and pinpoint what physics are important in particle acceleration [73].

Additional studies were also conducted using a custom simulated dataset of a Type III radio burst. Figure 16 shows a spectrum of the synthetic type III radio burst, with empirically realistic ramp up and ramp down times, as well as scattering of burst size over time. The right panel of this figure shows the correlated power measured by a radio interferometer for a diffuse Gaussian source with σ=10°. One can see how quickly this correlated power falls off for large antenna separations, and how necessary lunar enabled short separations are for full sky images.

Using this base model, we created a series of input images with Gaussian sources of growing sizes over time to simulate the scattering of the emission. It was found that the recovered power by *FarView* was in line with analytical predictions of correlated power as a function of frequency and baseline length, proving *FarView's* ability to characterize these scattering processes to an unprecedented degree. These studies show the necessity of having shorter antenna separations to observe highly scattered emissions, which can be more easily achieved on a stable location like the lunar surface.

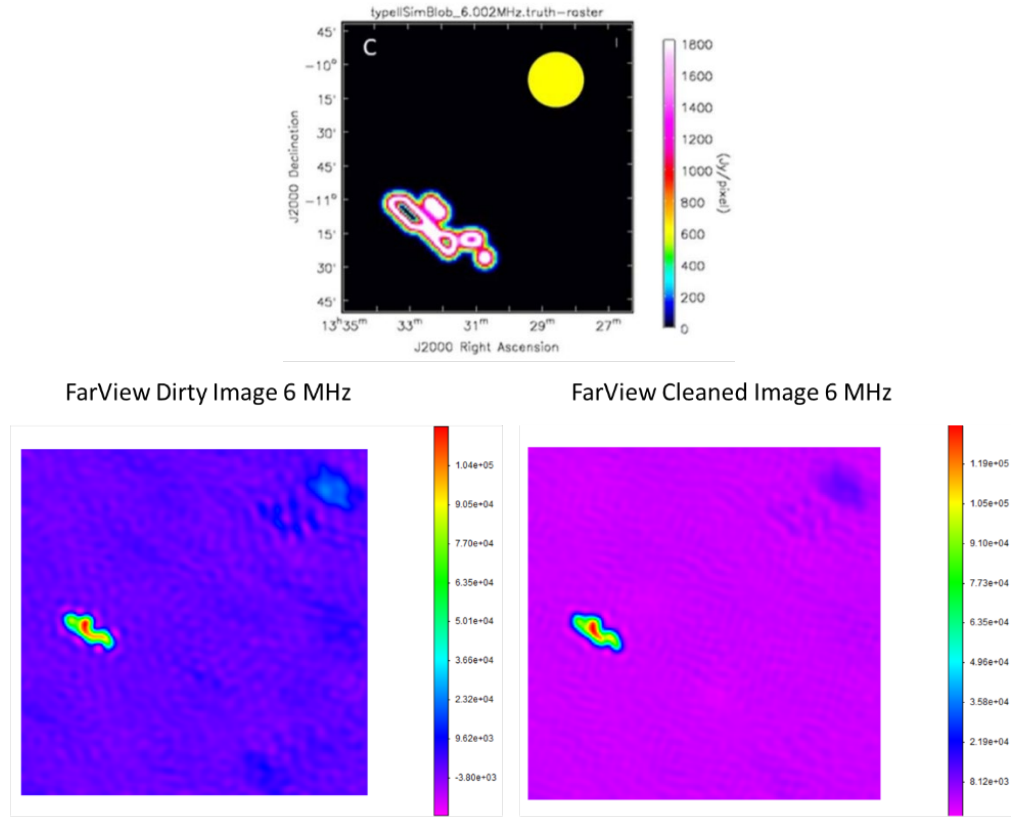

**Figure 15.** *Top:* Test input source distribution of a Type II burst at 6 MHz, roughly 5 solar radii out from the Sun. This is a brightness pattern from a simulated CME, with a hypothesized emission region upstream of the shocked region. *Lower left:* WSClean recovered dirty image from a FarView snapshot of the observed burst without deconvolution. *Lower right:* WSclean recovered clean image with several deconvolution cycles to reduce the effects of the dirty beam's sidelobes.

This work also demonstrates how FarView would directly contribute to decadal level science goals laid out in the ***Decadal Survey for Solar and Space Physics (Heliophysics) 2024-2033***[4]. This includes PSG 3b and 3c regarding understanding dominant fundamental energy conversion and transportation mechanisms that energize plasma and particles throughout the heliosphere and tracking the evolution of these solar eruptions. Additional studies could be done with the same instrumentation from the lunar surface to characterize fundamental plasma processes, contributing towards ***Moon to Mars objectives***. Relevant objectives include dust-plasma interactions (HS-

---

[4] https://www.nationalacademies.org/our-work/decadal-survey-for-solar-and-space-physics-heliophysics-2024-2033.

3LM) and understanding of magnetotail and pristine solar wind dynamics in the vicinity of the Moon (HS-4LM). Both objectives would be addressed by FarView measuring the variable quasi-thermal noise from the background electron densities found at the lunar surface in the 100-1000 kHz range.

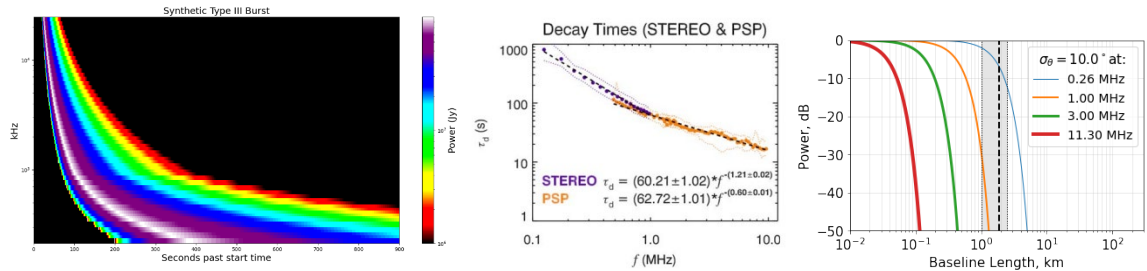

**Figure 16.** *Left:* Synthetic type III bust travelling at a speed of 0.3 c through the LeBlanc [37] electron density profile of the inner heliosphere. *Middle:* The ramp up and ramp down times that were applied according to observations from STEREO and PSP, taken from [74]. *Right:* Analytical predictions of correlated power for a 10-degree sigma Gaussian in the sky as a function of frequency and baseline length.

This work also demonstrates how FarView would directly contribute to decadal level science goals laid out in the ***Decadal Survey for Solar and Space Physics (Heliophysics) 2024-2033***[5]. This includes PSG 3b and 3c regarding understanding dominant fundamental energy conversion and transportation mechanisms that energize plasma and particles throughout the heliosphere and tracking the evolution of these solar eruptions. Additional studies could be done with the same instrumentation from the lunar surface to characterize fundamental plasma processes, contributing towards ***Moon to Mars objectives***. Relevant objectives include dust-plasma interactions (HS-3LM) and understanding of magnetotail and pristine solar wind dynamics in the vicinity of the Moon (HS-4LM). Both objectives would be addressed by FarView measuring the variable quasi-thermal noise from the background electron densities found at the lunar surface in the 100-1000 kHz range.

### 5.1.4 Summary of Observations of Solar Radio Bursts with FarView

The *FarView* concept array has been effectively advanced in this Phase II study to demonstrate the exciting prospects of heliophysics and space weather research with *FarView*. The simulation software has been upgraded to be more flexible in terms of central location, enabling studies of the effect of lunar latitude on *FarView's* observations. With this updated design, the simulation software has been used to prove out the capabilities of the array to achieve key science goals in heliophysics. This includes making detailed images of low frequency emissions from Type II and III solar radio bursts. This will help us understand new aspects of the solar energetic particle acceleration process around coronal mass ejections and solar flares that are only visible with radio interferometric observations on the Moon.

### 5.2 Galactic and Stellar Astrophysics with *FarView*

*FarView will open a novel new window on Galactic and stellar astrophysics. A central objective is cosmic ray tomography, exploiting free-free absorption along the Galactic plane to reconstruct the first three-dimensional map of energetic electron distribution in the Milky Way's interstellar medium. The observations will provide essential "ground truth" for models of*

---
[5] https://www.nationalacademies.org/our-work/decadal-survey-for-solar-and-space-physics-heliophysics-2024-2033.

*turbulence, magnetic dynamics, and Galactic flows. We describe in this section how this breakthrough addresses critical **Astro2020** priorities on feedback, magnetic fields, and ISM structure—key drivers of galaxy evolution. FarView will also revolutionize our understanding of stellar space weather, a dominant factor in planetary habitability yet poorly characterized beyond the Sun. It will detect long-wavelength Type II and III radio bursts—signatures of shocks and electron beams from stellar eruptions—from all solar-type stars within 100 parsecs. These measurements will probe magnetic field geometries, CME occurrence, and particle acceleration, constraining stellar dynamo models and coronal heating mechanisms. By systematically surveying bursts across stellar masses, ages, and rotation rates, FarView will trace magnetic evolution, quantifying its impact on exoplanet atmospheres and biosignature false positives.*

The **Astro2020 Decadal survey** [44] identified emerging opportunities for high impact advances over this decade and beyond to resolve long standing gaps in our understanding of cosmic ecosystems and the drivers of galaxy growth, the physics of stars, and star-planet connections. The survey identified driving questions, including: How do gas, metals, and dust flow into, through, and out of galaxies? How do star-forming structures arise from, and interact with, the diffuse interstellar medium? What regulates the structures and motions within molecular clouds? What are the most extreme stars and stellar populations? What would stars look like if we view them like we do the Sun? How do the Sun and other stars create space weather?

The decameter and kilometric radio bands (>10-meter wavelengths, <30 MHz frequencies) provide unique windows to address these critical questions. However, as noted in the previous sections, such ultra-long radio wavelengths are inaccessible from the surface of the Earth due to Earth's ionosphere that becomes opaque for wavelengths greater than 15 meters and distorts propagation for wavelengths as short as 1 meter. Decameter and longer wavelengths are accessible from space yet still contaminated by anthropogenic transmitters that leak through Earth's ionosphere, strong auroral kilometric radiation due to the solar wind interacting with Earth's magnetosphere, and solar radio bursts. Locating radio telescopes on the far side of the Moon addresses these remaining hurdles, providing a clean radio environment for astronomical students that is unique in the inner solar system.

Here we describe two science objectives themed around Galactic and stellar astrophysics that are enabled by the *FarView* lunar farside radio array.

### 5.2.1 Cosmic Ray Tomography

Mapping free-free absorption from electrons in ionized regions along the Galactic plane at decameter wavelengths offers a transformative opportunity to create the first three-dimensional tomographic map of the Milky Way's distribution of cosmic rays in the interstellar medium (ISM). Observations from *FarView* would penetrate a long-standing observational gap, providing "ground truth" for models of feedback and structure formation in the ISM.

As noted in the **Astro2020 Decadal Survey** [44], "massive stars and their supernovae are sources of mass, energy and momentum, which emerge in the form of fast-moving shock-heated gas, and relativistic cosmic ray particles as well as photons." (p. 74) The **Astro2020 Decadal Survey** further highlighted that "In addition to understanding the energy, mass, and momentum that stars supply to their surroundings, determining how these winds, radiation, and supernovae interact with the surrounding gas on different scales in the interstellar medium is equally important to untangling their interplay." (p. 69)

Cosmic rays of energetic electrons regulate energy exchange within a galaxy and are critical to understanding galactic ecosystems [45]. Supernova remnants lose 10-50% of their energy through cosmic rays, slowing their expansion times, increasing their momentum transfer into the ISM, and diffusing their energy over kiloparsec scales. Thus, cosmic rays drive heating and ionization, maintaining minimum temperatures even in dense interstellar clouds and regulating the chemistry of the clouds. They maintain turbulence in the ISM and drive Galactic flows. *Astro2020* emphasized this critical role and the limitations of our present knowledge, stating "The impact of cosmic rays is one of the largest uncertainties in understanding feedback in galaxy formation." (p. 70)

Cosmic rays interact with Galactic magnetic fields, governing the diffusion of the energetic particles away from their sources, and generating synchrotron radio emission that provides a critical window into both the electron density of the ISM and magnetic fields. *Astro2020* noted "The primary uncertainty is how cosmic rays are scattered by small-scale fluctuations in the magnetic field, which sets whether cosmic rays can escape a region or whether their pressure builds up to the point where it can drive an outflow", stating further that "fluctuations in the galactic magnetic field are a key ingredient in understanding how galaxies drive winds on scales of tens of kiloparsecs", and concluding that "large scale magnetic field properties or distant supernovae can affect the formation of pre-stellar cores." (p. 70)

The Galactic radio spectrum increases with wavelength following approximately a power-law spectrum driven by the synchrotron emission from cosmic rays. Around 100 m wavelengths (3 MHz) the total sky power begins to roll over at longer wavelengths due to free-free absorption in partially ionized gas clouds, called HII regions. These HII regions of high electron density eventually become opaque to radio waves at long wavelengths. For ionized clouds at known distances, this effect can be used to separate the contribution to the synchrotron emissivity from behind the cloud compared to in front. With long wavelength radio sky maps and sufficiently large numbers of identified HII regions, the three-dimensional distribution of electron density can be reconstructed [45,46,47,48,49, 50] by combining the radio emissivity measurements with magnetic field observations from polarimetry at higher frequencies or gamma-ray observations to break the degeneracy between cosmic ray density and magnetic field strength in synchrotron emissivity. Figure 17 illustrates these effects and the potential for decameter radio observations to uniquely yield three-dimensional, tomographic mapping of the electron density in the Milky Way. Observations at higher frequencies (10s to 100s MHz) already see individual HII regions increasing in optical depth. Extension of these observations to 0.1 to 10 MHz by lunar radio telescopes will enable mapping across the entire Galaxy.

Tomographic imaging derived from long wavelength radio observations will uniquely contribute to overcoming the current limitation highlighted in *Astro2020* that "our view of the Milky Way's ISM has been hampered by seeing it in projection." (p. 315). Ultimately, lunar-based observations will address the Astro2020 objective of "delineating the 3D structure of the gas, dust, and magnetic field, providing a new and dramatically detailed view of the dynamic ISM" (p. 314) and enable new tests for models of Galactic feedback and evolution.

### *5.2.2 Instrument requirements for cosmic ray tomography*

The first measurements of the evolution of ISM opacity due to free-free absorption at ultra-long wavelengths will be performed with a single antenna instrument on the lunar farside, *LuSEE-Night*, observing over a full lunar month (30 Earth days), to create a drift scan light curve of the variation in Galactic brightness temperature as the sky moves overhead [52]. Acquiring drift scans from 0.1 to 30 MHz would show significant changes as a function of wavelength in the dynamic range of the light curve between periods when the inner Galactic plane–rich in HII regions–is overhead compared to periods when the outer plane or high Galactic latitudes are overhead. Figure 18 illustrates a notional example following the style of observations acquired by the EDGES ground-based spectrometer [66, 67]. These simple observations are not sensitivity-limited because the Galactic emission fills the instrument beam and dominates over all other noise sources.

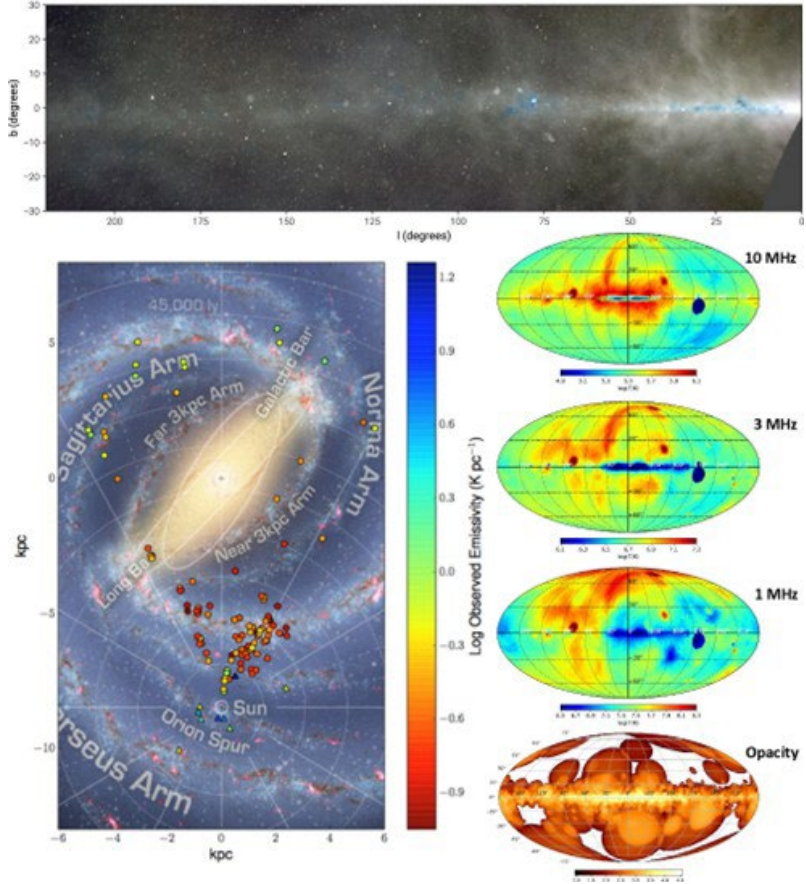

**Figure 17**. *(top)* Observed increase in optical depth at low frequencies (36-73 MHz) due to free-free absorption in the Galactic plane (see blue regions in the top three-color plot where lowest frequency is in red and highest in blue [51]. *(left)* Polderman et al. [45] report fits changes in radio emissivity along the sight lines to a sample of known HII regions. *(right)* Simulated radio emissivity at three frequencies showing increasing absorption in the Galactic plane and modeled integrated optical depth at 1 MHz [47].

Compact interferometric arrays with maximum baseline lengths of order 1-10 km–such as those envisioned for Dark Ages 21-cm cosmological observations using *FarView* [55, 58]–would enable degree-scale regions along the inner Galactic plane to be resolved, adding further spatial information for constraining cosmic ray and magnetic field models. Figure 19 shows simulated images that suggest such an array is sufficient to distinguish areas along the Galactic plane that are rich in HII regions, corresponding to the high absorption areas in the ultra-long wavelength sky model of [47].

Ultimately, larger interferometers with 100 km baselines and hundreds of antennas, yielding high-quality arcminute imaging matched to typical HII regions scales, are needed to unlock the full potential of cosmic ray tomography. These instruments could identify individual HII regions for cross-referencing with other studies providing distances to separate the synchrotron emissivity

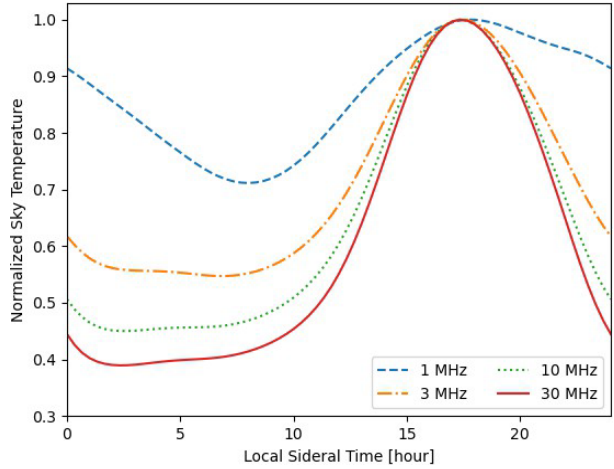

**Figure 18**. Representative illustration of the variation in average Galactic brightness temperature over a full lunar day. The plot shows simulated peak-normalized drift scans for a single dipole antenna on Earth (at -27 deg latitude) over a 24-hour period. Earth's ionospheric absorption is omitted to match the lunar environment and the ultra-long wavelength sky model in [47] is used to capture the effects of free-free absorption along the Galactic plane. The absorption is evident as a reduction in the dynamic range at lower frequencies between LSTs when the Galactic inner plane is overhead (peaking at LST 18h) and other LSTs when high Galactic latitudes or the outer plane are overhead. Light curves created over a full lunar day (30 Earth days) by an instrument on the Moon would enable model fits to measure the average free-free absorption as a function of frequency.

contribution along the line of sight and magnetic field properties.

### 5.2.3 Space Weather in Stellar Systems

Stellar systems are critical drivers of galactic evolution and cosmic ecosystems, as well as hosts to planetary systems. *Astro2020* emphasizes "The Sun and other stars affect their environments in numerous ways, from the interaction of stellar winds, flares, coronal mass ejections (CMEs) and other forms of mass loss with surrounding disks, planetary bodies, stellar companions, and the interstellar medium, to the creation of planetary nebulae and supernova remnants and the end stages of stars." (p. 339) It further highlights that "Identifying the physical processes involved in these interactions informs a broad range of current astrophysical problems, from stellar feedback in galaxy evolution, to the formation and retention of atmospheres on planets."

Many of these stellar phenomena are driven by stellar magnetism, rotation, asymmetries, and surface processes, properties that have been difficult to study for stars other than our Sun. Indeed, *Astro2020* assesses that, in astrophysics, "stars are typically treated observationally as featureless points of light and theoretically as spherically symmetric objects. However, observations of our own Sun demonstrate that stellar surfaces are complex and dynamic, and sometimes not even spherical. Spots, flares, tidal distortions, mass-loss, rotation, and internal convection all break the spherical symmetry of stars. Observations in the past decade have revealed the pervasiveness of these effects, which have confounded our ability to understand the fundamental properties of stars and their surroundings".

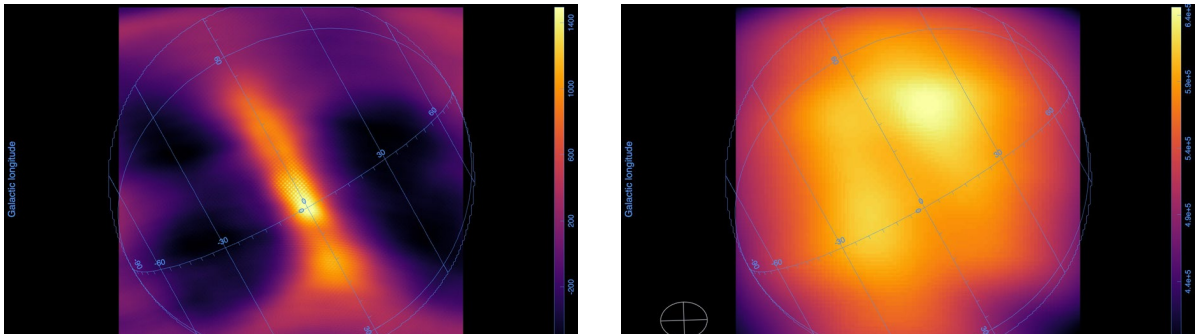

**Figure 19.** Representative illustration of simulated "dirty" (not deconvolved) sky images using the ultra-long wavelength sky model in [47] and an array of 128 dipole antennas matching the layout of the Murchison Widefield Array (MWA) ground-based telescope with a compact core surrounded by outriggers reaching 3 km baselines. The left panel shows the simulated sky at 30 MHz and the right panel at 1 MHz. The inner Galactic plane is clearly visible in strong emission as a diagonal band in the left panel but decreases to a darker band relative to the surrounding sky in the 1 MHz image due to free-free absorption. Image credit: D. Jacobs, J. Pober, and J. Bowman.

Of the needed new probes, long wavelength radio observations provide a powerful new window into stellar magnetism, particle acceleration, and mass loss. *Astro2020* highlights that "The growing interest in exoplanet atmospheres and potential habitability mandates a better theoretical understanding of stellar magnetism and its effects throughout a star's system."

On the Sun, CMEs and solar energetic particle events are typically accompanied by solar flares spanning from X-ray to radio wavelengths. Solar radio bursts are the most intense sources of astronomical radio emission detected from Earth, reaching $10^{12}$ Jy. Type II solar radio bursts are slow-drifting radio emissions at meter-to-kilometer wavelengths caused by shock waves accompanying electron production from solar eruptions. They are key indicators of space weather events. The most energetic CMEs are associated with broadband Type II bursts [53]. Another key category of bursts, Type III, are fast-drifting radio emissions observed from centimeters to 10s of kilometers wavelengths. They are among the most common and dynamic solar radio phenomena, associated with electron beams accelerated during solar flares that excite plasma oscillations in the solar corona and interplanetary space, emitting radio waves at the local plasma frequency and its harmonics [54].

Ultra-long wavelength lunar radio telescopes will be able to detect the equivalent of Type II and III solar bursts at decameter and longer wavelengths for the first time from all solar-type stars within 10 parsecs for the FARSIDE array [85] and over 100 parsecs with *FarView*. Expanding access to observations of stellar radio flares equivalent to long wavelength solar Type II and III radio bursts will directly address the decadal survey's key question G-Q4: How do the Sun and other stars create space weather?

Long wavelength radio bursts trace the propagation of shocks and electron beams driven by magnetic reconnection through stellar coronae. Detecting stellar radio bursts reveals how magnetic activity varies across different types of stars, offering insights into stellar dynamos, coronal mass ejections, and space weather environments that we cannot access by studying the Sun alone. Burst frequency drift rates, polarization properties, and harmonic structure encode information about magnetic field strength and geometry. This could help address the *Astro2020* objective to understand the "diversity of coronal heating mechanisms", which may be "far more complex than currently appreciated." Type III bursts arise from flare-accelerated electron beams escaping along open field lines into interplanetary space. Their frequency drift and intensity provide diagnostics of plasma density gradients and acceleration mechanisms.

Observing these bursts across a range of stellar types may constrain dynamo models and reveal how magnetic topologies evolve with stellar mass and rotation, and whether electron acceleration is universal or varies with stellar parameters. By observing stars at different evolutionary stages, low-frequency radio bursts can trace the decay of magnetic activity over time. This complements solar studies, which are inherently limited to a single age and activity level. *Astro2020* emphasizes "current observations tend to be biased toward the nearest and/or most active stars. In the next decade, we need to characterize transient energetic phenomena systematically, as a function of stellar type, mass, age, metallicity, and rotation rate, and explore these phenomena across the electromagnetic spectrum. Direct observation of stellar winds and CMEs will help constrain these effects."

Type II bursts are direct signatures of CME-driven shocks. While CMEs are well-characterized in the solar context, their prevalence and properties in other stars remain uncertain. Detecting Type II-like bursts from M-dwarfs and solar analogs will establish whether CMEs are common and assess their role in shaping planetary atmospheres and habitability. M-dwarfs exhibit extreme magnetic activity, including frequent flares and coherent bursts [56]. Some M dwarfs exhibit auroral-like radio bursts, analogous to planetary aurorae. These are likely driven by magnetospheric currents rather than flares and may persist even in quiescent states. This behavior is especially pronounced in ultracool dwarfs, which blur the line between stars and brown dwarfs. Recent detections of coherent bursts from YZ Ceti that are in phase with the orbit of its closest terrestrial planet suggest that magnetic star–planet interactions may be common in these systems [57]. Such phenomena are absent in the Sun, underscoring the need for broader observational baselines.

M-dwarfs host the majority of exoplanets in the Milky Way and are expected to each produce extreme flares on monthly cadences. The *Astro2020 Decadal Survey* further highlights "Stellar flares on active stars extending to the lowest stellar masses, observed from the radio to the X-ray, can have high-energy luminosities up to five orders of magnitude larger than flares on the Sun." These bursts would have significant impact on the exoplanets orbiting the host stars, causing atmospheric mass loss that could limit habitability and create atmospheric chemistries that could be confused with biosignatures. Scaling models predict these bright bursts will peak at wavelengths longer than 30 meters (below 10 MHz) and, thus, are not visible to ground based radio telescopes. Characterizing these events with lunar radio telescopes is critical for habitability analysis.

### *5.2.4 Instrument requirements for stellar space weather*

Solar CMEs and SEP events are accompanied by radio bursts at low frequencies, i.e., Type II radio bursts and a subset of Type III radio bursts. The emission is produced at the fundamental and first harmonic of the plasma frequency and provides a diagnostic of the density and velocity (few 100 to >1000 km/s) near the shock front, while the flux density of the burst depends sensitively on the properties of the shock and solar wind [89]. Ground-based radio astronomy can trace such events only to a heliocentric distance of a few solar radii, whereas by probing lower frequencies (<30 MHz) lunar-based radio telescopes can trace the propagation of shocks out to the Earth and beyond, which is particularly relevant for characterizing geoeffective CMEs and SEP events. The detection of equivalent interplanetary Type II and III events from stars other than the Sun is one of the goals of the *FarView* array.

*Farview* will detect the equivalent of the brightest Type II and Type III bursts out to 10 pc at frequencies below a few MHz. By imaging >10,000 square degrees every 60 seconds, it would monitor a sample of solar-type stars simultaneously, searching for large CMEs and associated SEP events. In addition, the panoptic capabilities of the beam forming mode of *FarView* where we could have multiple pointing for longer durations on dedicated targets, would help with accessing burst rates [90]. For the Alpha Cen system, with two solar-type stars and a late M dwarf, it would probe down to the equivalent of $10^{-14}$ W m$^{-2}$ Hz$^{-1}$ at 1 AU, a luminosity at which solar radio bursts are frequently detected at the lowest frequencies available to *FarView* [91]. The nearby young active solar-type star, Epsilon Eridani (spectral type K2), is another priority target. For the case of M dwarfs, *FarView* would be able to detect Type II bursts formed at the distance where super-Alfvénic shocks should be possible for M dwarfs and directly investigate whether the relationship observed between solar flares and CMEs extends to M dwarfs.

Since we are interested in the lowest frequencies of the FarView bandwidth, it is important to ensure that the antennas will be sky-noise dominated at the relevant frequencies. Figure 20 shows the sky noise dominance factor for the *FarView* dipoles, considering the impedance mismatch and the fraction of the beam above the horizon. Over the frequencies of interest, between 5 and 20 MHz, the sky noise dominance factor varies between 3× and 100× for the *FarView* dipoles making it suitable for this science case.

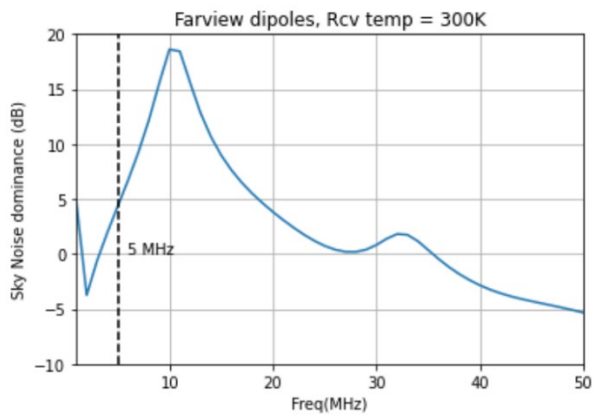

**Figure 20**: Sky noise dominance factor of the 10-m *FarView* dipoles in dB. The dipole was simulated by placing it on a realistic lunar regolith with layered structure and electrical properties from the Lunar Source Handbook. The vertical line at 5 MHz indicates the lower end of the *FarView* bandwidth. For the stellar space weather science case, we are interested in frequencies <20 MHz, where the antennas are sky-noise dominated (>5 dB). This calculation was performed considering the beam above the horizon and the antenna's mismatch factor.

### 5.3  Identifying Habitable Exoplanets with Farview

*The search for habitable worlds beyond Earth requires not only planetary atmospheric measurements with facilities like JWST, the Roman Space Telescope, and the Habitable Worlds Observatory, but also detection of magnetospheres, which are essential for shielding atmospheres from the intense stellar activity of M-dwarfs—the most common exoplanet hosts. In this section, we describe how low frequency radio bursts uniquely reveal magnetic fields. On Earth, our ionosphere blocks these signals, making the lunar farside the only viable site for detection. Farview's 100,000 dipole elements can achieve sub-milliJansky sensitivity in hours and microJansky sensitivity over long integrations below 50 MHz. This is sufficient to detect auroral emissions from exoplanets around M-dwarfs within 5 parsecs. During heightened stellar activity, flux densities may increase by factors of ~1000, allowing rapid detection within a few hours and for more distant planetary systems. The sensitivity and field of view of FarView will enable the first statistical surveys of exoplanet magnetospheres, distinguishing planetary signals from stellar bursts through polarization and rotational modulation.*

The search for habitable conditions beyond Earth is a top priority in astrophysics. The discovery of habitable exoplanets beyond our solar system will require a suite of instruments providing long-term monitoring for detection (e.g. with space and ground-based radial velocity observations), spectroscopic characterization of atmospheric and surface properties, and eventually deep chronograph-aided observations from e.g. JWST, Roman Space Telescope, and the Habitable Worlds Observatory (HWO). Detection of exoplanet magnetospheres is necessary to identify the most promising targets for follow-up characterization of biosignatures with these assets, and to provide an ensemble of objects for studies of magnetospheric conditions and atmospheric composition. Strong planetary magnetospheres are critical for retaining atmospheres needed for life. As noted in the **Planetary Science Decadal Survey**, "M-dwarf stars which house exoplanets often exhibit extreme levels of stellar activity, which may adversely impact the habitability of orbiting planets." Enhanced radiative output during stellar flares has been shown to expand and expel atmospheres. Fast, dense winds, especially from young stars and low-mass M-dwarf stars (the most common stellar type), can compress magnetospheres and expose atmospheres to mass loss [81, 82], particularly during coronal mass ejections (CMEs). This has been tested in our Solar System. MAVEN found evidence that ion loss from solar winds strongly depleted the early Martian atmosphere [83].

Only observations of low-frequency radio emission will distinguish exoplanet magnetospheres [87]. As identified by the **Planetary Science Decadal** recommendations, "Excited aurorae in the waveband spanning Radio to X-ray spectrum are a window into the workings of these magnetospheres." Exoplanet radio bursts indicate the presence and strength of a magnetic field. For example, all magnetized planets in our solar system produce coherent radio emission by auroral processes powered by electron cyclotron maser instability, typically through magnetic reconnection between the planetary field and the field carried by the solar wind, as is the case with Earth, Saturn, Uranus, and Neptune. Jupiter's primary radio emission is driven by the solar wind compressing its co-rotating plasma sheath and by interactions with the magnetic fields of its moons. In all cases, the characteristic frequency of radio emission is proportional to the electron cyclotron frequency, which is determined by the magnetic field strength: ν ≈ 2.8×B(Gauss) MHz. Jupiter's strong ≈14 G magnetic field produces radio emission with peak frequency of 40 MHz. All other magnetized planets in our solar system have magnetic fields <2 G, making their peak emission frequencies below 6 MHz. Earth's aurora kilometric radiation (AKR) peaks around ~0.3 MHz.

Detecting terrestrial-like exoplanet radio emission requires a large collecting area on a space-based observatory to avoid Earth's opaque ionosphere (below 10 MHz) and AKR. The non-polar lunar farside is the only suitable location in the inner solar system since it is shielded from Earth's emissions [84]. Lunar radio interferometric arrays, such as the FARSIDE pathfinder and the larger *Farview* [85, 58], address this and other high-priority Astro2020 decadal science objectives. These lunar radio telescopes will survey large areas of the sky simultaneously to detect radio bursts by exoplanets in the solar neighborhood. FARSIDE and *Farview* are designed to image up to 10,000 square degrees simultaneously. These large fields will include 2000 stellar systems within 25 parsecs.

Stellar winds and mass ejections needed to feed the auroral activity can be sporadic, necessitating large fields of view and continuous surveying. Exoplanet emission will be identifiable by its circular polarization, distinguishable from bursts of the host star by rotational modulation and not limited by classical source confusion because the interstellar medium plasma is optically thick at very low frequencies.

### 5.3.1 FarView Sensitivity

We use the radiometric sensitivity equation to estimate the weakest signals that the 100,000 element *Farview* array can detect:

$$\sigma = \frac{2k_B(T_{sky} + T_{rcv})}{\eta_A \sqrt{n_p N(N-1)\Delta\nu \ t_{int}}} * 10^{-26} \text{Jy}$$

where $\eta_A = \lambda^2/4\pi$ is the collecting area for an ideal dipole and N is the number of dipole antennas. Simple dipoles of 10-m length are what will serve as the array elements for the *Farview* telescope. For $T_{sky}$, we use the reference sky temperature of 5000 K at 50 MHz and a spectral index of -2.5 to obtain the relevant sky temperatures in the *FarView* frequency range (see also Section 7). At these frequencies, $T_{rcv} \ll T_{sky}$. In a mere 2.5 hours of integration, *Farview* can achieve sensitivities of sub-milliJansky (as seen in the blue curves of Figure 21). And in 2500 hours, *Farview* can detect sources at the micro-Jansky level.

In all the sensitivity curves shown in Figure 21, we see that at frequencies <7 MHz the sensitivity gets worse because at those wavelengths the effective area ($\lambda^2/4$) of the dipoles becomes greater than their physical area. This means that the dipole-dipole distance in the short baselines is less than a wavelength, implying that the effective number of dipoles to be considered in the sensitivity calculation is reduced. We take this into account by recalculating the sensitivity assuming that the number of antennas = total area/effective area of the dipoles. For example, the effective number of antennas in the core at 7 MHz is slightly reduced to 80,536 from 82,944.

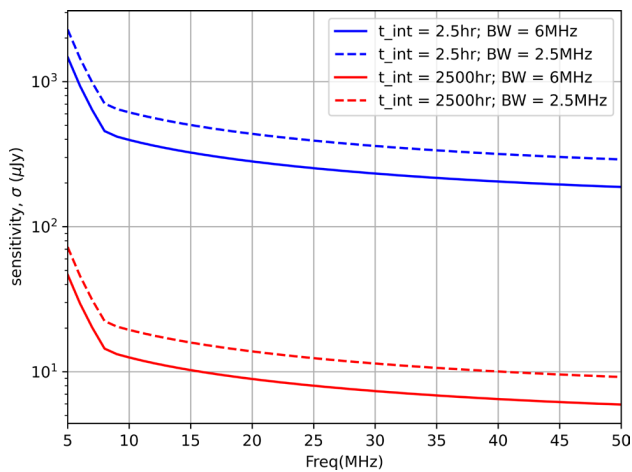

**Figure 21:** Sensitivity of the 100,000-element *FarView* telescope versus frequency for a few different observation modes: blue corresponds to an integration time of 2.5 hours, and red is plotted for a longer integration period of 2500 hours. For each integration period, the solid lines correspond to a 6 MHz bandwidth and the dashed lines to a 2.5 MHz bandwidth.

### 5.3.2 Exoplanet Detectability by FarView

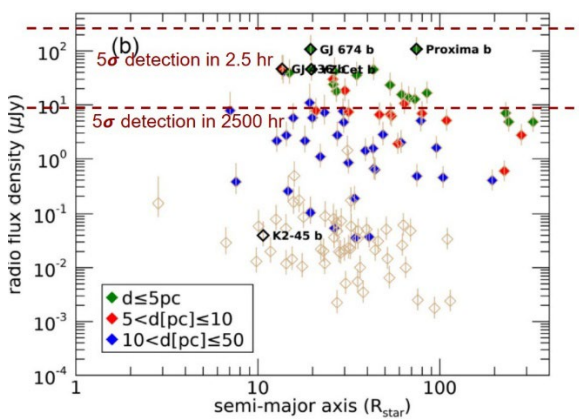

**Figure 22**: Radio flux densities of all the exoplanets discussed in [86] assuming magnetic strengths of 1 G, 0.1 G, and 10 G for the symbols and upper and lower error bars, respectively. Overplotted in red dashed lines are the 5$\sigma$ sensitivity of *Farview* for two values of integration times: 2.5 hr and 2500 hr.

Given the sensitivity that is achieved with Farview, we look at the detectability of radio auroral emission from various exoplanets discussed in [86]. With 2500 hr integration, Farview's sensitivity is ≈7 µJy at 27 MHz. Plotting this against the expected radio flux densities of known exoplanets in Figure 22, we see that *Farview* can achieve a 5$\sigma$ detection of radio emissions from all known exoplanets (with semi-major axis < 100pc) around M-dwarfs within 5 pc. To obtain radio emission from the exoplanets around 27 MHz, the assumed planetary magnetic field strength is $B$=10 G (~20× that of Earth's) since $\nu \approx 2.8 \times B$ MHz. All the calculations shown in Figure 22 are during the quiet solar wind period. During high solar winds/CMEs, we can expect the emitted flux to increase by ~1000, implying *Farview* can detect these bursting auroral emissions in the 2.5-hour integration period.

## 6.0 The *FarView* Radio Interferometer versus a Single Dish Radio Telescope for 21-cm Dark Ages Cosmology on the Moon

There are two general classes of radio telescopes to consider for a Cosmic Dark Ages cosmology mission: single dishes and interferometers. The ground-based 21-cm cosmology community has largely focused on developing interferometers, but it is fair to evaluate if this is also the right choice for the Dark Ages on the Moon. In assessing the tradeoffs between the two approaches, there are three key aspects to consider: (1) the range of spatial scales probed; (2) the susceptibility to and ability to mitigate systematic errors; and (3) the engineering constraints given the lunar far side location.

The biggest limitation of a single dish experiment is generally the angular resolution and, therefore, the range of spatial scales that it can map. But the 21-cm signal has most of its power on large scales and generally does not require high resolution to probe the features of interest (although many theories do predict novel physics showing up on very small scales [e.g., 76, 77]). However, with 21-cm cosmology, one must also recall that experiments probe a three-dimensional volume, and even with poor angular resolution, very small scales can be accessed along the line-of-sight simply by increasing the spectrometer resolution. Another argument that may seem in-favor of interferometers is the need to survey large volumes, which they can do more efficiently given the very large field of views enabled by small individual antenna elements. However, for a *detection* experiment, it is generally favorable to focus on a narrow area of sky. Precision cosmological measurements may require very large survey volumes to reduce sample variance, but that is likely the realm of a more ambitious future experiment. Given these considerations, neither a single dish nor an interferometer has a particularly strong advantage when it comes to intrinsic sensitivity.

However, the two approaches present a stark difference when it comes to systematic errors. Both have the potential for significant systematics with complicated frequency structure, e.g., from long wavelength standing waves in a single dish and, e.g., from mutual coupling between elements of an interferometer. But the most significant source of systematic error for 21-cm experiments is foreground emission and the daunting foreground-to-signal ratio of >$10^5$ during the Dark Ages necessitates aggressive foreground removal. Accurate foreground modeling is therefore critical for a Dark Ages experiment and sets a requirement of high angular resolution such as can only be achieved with the long baselines of an interferometer. Future studies are necessary to better quantify the foreground modeling requirements for both kinds of experiments, but it is clear interferometers have a very significant advantage in characterizing the small-scale structure of the foreground sources.

Third, the engineering challenges of *FarView* versus a single dish telescope are very different. For *FarView,* we have recommended that simple dipole antennas can be deployed on the lunar surface using teleoperated rovers in a relatively straightforward fashion [58]. We have proposed MRE technology, successfully demonstrated in the laboratory, to fabricate the antennas in-situ using aluminum oxide within lunar regolith as feedstock. A single lander, such as the Blue Moon from Blue Origin, can deliver metric tons to the lunar surface including preamps, an MRE reactor, and a rover to build a substantial cluster of antennas. We can begin doing science immediately with this initial (sub)array and then grow its capability with additional antennas. In contrast, the construction of a single dish telescope on the Moon, such as the Lunar Crater Radio Telescope (LCRT) [75], is substantially more challenging, involving multiple landers, rovers, and construction techniques that have not been demonstrated off-world as of yet. Science cannot begin until the entire dish is operational. The relative simplicity of a radio interferometer coupled with its superior resolution make it a clear choice for Dark Ages cosmology on the lunar far side.

## 7.0 The Urgency of the *FarView* Radio Array – Unintended Electromagnetic Radiation (UEMR)

Recently, Starlink satellites in Earth orbit have been detected by the ground-based low frequency array LOFAR in Europe and by the OVRO LWA in California as unintended electromagnetic radiation (UEMR), likely caused by RF leakage from onboard power supplies and electronics. Bassa et al. [78] reported that "the spectral power flux density of the broadband UEMR [from Starlink satellites at an average altitude of 500 km] varies from satellite to satellite, with values ranging from 15 to 1300 Janskys (Jy), between 56 and 66 MHz, where 1 Jy = $10^{-26}$ W/Hz/m$^2$. We have scaled this signal for hypothetical satellites at different orbits around the Moon to calculate the potential UEMR experienced on the lunar surface by telescopes such as *FarView*.

To determine the confidence level of the detection (RMS noise), we have assumed that the primary noise source is radiometer noise given by the Radiometer Equation:

$$\sigma = T_{sys}/(Bt)^{1/2}$$

where $\sigma$ is the RMS uncertainty in the noise temperature measurement, $T_{sys}$ is the noise temperature of the radiometric observation, $B$ is the bandwidth, and $t$ is the integration time. The dipole antennas that we use for *FarView* are electrically short yet are efficient enough to be sky-noise dominated [55] (i.e., much less than internal noise from the radio receiver). So, the system or sky temperature ($T_{sys} \approx T_{sky}$) is dominated by the Galaxy brightness temperature (produced by synchrotron emission above ~5 MHz),

$$T_{sky} \sim 5000 \text{ K } (\frac{\nu}{50 \text{ MHz}})^{-2.5}.$$

The RMS flux density sensitivity of *FarView* depends upon the frequency bandwidth, the integration time, the system temperature, and the effective area of the array. The effective collecting area for the array is $\eta_A \times \sqrt{N(N-1)}$ where $\eta_A$ is the effective area for a single dipole antenna[6] and N is the number of dipoles. The resulting *FarView* RMS sensitivity, or minimum detectable flux density ($S_{min}=\sigma$), is given by

$$\sigma = \frac{2kT_{sky}}{\eta_A\sqrt{N(N-1)Bt}}.$$

---

[6] $\eta_A = \lambda^2/4\pi$ = 32 m$^2$ per dipole or ~3.2 km$^2$ for the full array collecting area at 15 MHz, and 4.4 m$^2$ at 40 MHz.

After only 1 minute of integration with $N=10^5$, $B=0.5f$ and with $f$ as the center frequency of the band, the RMS sensitivity for broadband total power measurements is ~4.1 mJy at 15 MHz and ~1.6 mJy at 40 MHz.

To calculate the signal-to-noise, *S/N*, of *FarView* required to detect satellites in cis-lunar space, we assume that all 100,000 antennas are operating and that the observed flux density falls off with distance (r) as $1/r^2$. As an example, let us suppose that the detected power of a Starlink-like target has a source flux density of S~100 Jy (mid-way in range detected by Starlink at an altitude of 500 km above Earth). Furthermore, we observe this satellite with *FarView* at 1000 km above the lunar surface at 40 MHz with a 20 MHz bandwidth and use an integration time of 1 hour ($\sigma$=0.2 mJy). The result of this observation is a highly significant detection with

S/N = S/ $\sigma$ = 100 Jy×(500/1000)²/(1.6 ×10⁻³ Jy × $\sqrt{1\ min/\ 60\ min}$ ) = 25 Jy/(2×10⁻⁴ Jy)

= 121,000.

Placing that satellite at different altitudes above the Moon with 1 hour integration times at 40 MHz yields detection signal-to-noise ratios of:

- For 1,000 km, the observed flux density is 25 Jy for a *S/N*= 121,000.

- For 10,000 km, the observed flux density is 250 mJy or *S/N* = 1210.

- For 70,000 km (apolune for distant retrograde orbit used by the Lunar Gateway), the observed flux density is 5 mJy or *S/N*= 25.

If we assume the higher, maximum published observed flux density of 1300 Jy for a Starlink satellite at 500-m altitude above Earth, then all the *S/N* ratios above scale by a factor of 13. Even a satellite at a distance of ~1 million km from the Moon (e.g., distance of JWST) observed with *FarView* can be detected with a flux density of 325 microJy for an integration time of 24 hours ($\sigma$ = 41 µJy) or S/N=8.

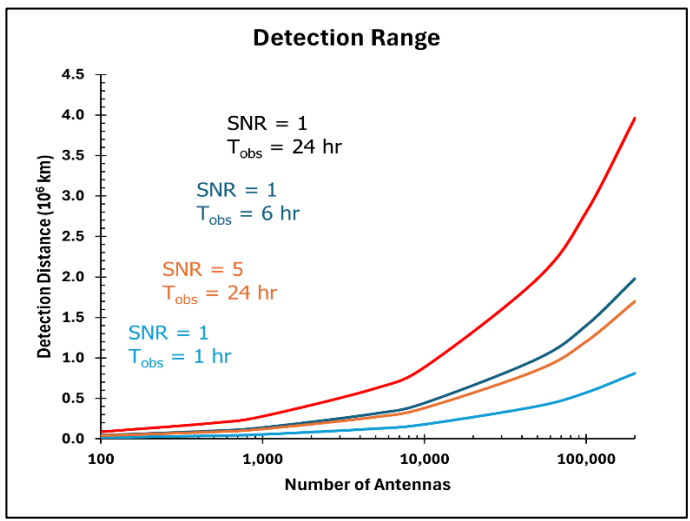

**Figure 23.** Detection distance from the Moon for Starlink-like satellites in cis-lunar space using *FarView*. We assume observations with 100,000 dipoles distributed over 200 km², 24 hour integration period, center frequency of 40 MHz, bandwidth of 20 MHz. and emitter flux density of 1300 Jy at an altitude of 500-m.

In Figure 23, we illustrate the distance from the Moon that Starlink-like satellite UEMR, visible from the far side, can be detected with *FarView*. From this analysis, we conclude that if there are satellites with comparable radio frequency leakage from electronics like that of Starlink placed in lunar orbit, the radio-quiet of the far side at low frequencies due to UEMR is in great danger! Therefore, it will be important to build *FarView* and their prototypes as quickly as possible while the Moon has minimal UEMR.

## 8.0 Data Rates and On-site Computational Load for *FarView*

*FarView's design must confront a fundamental operational challenge: processing vast volumes of data and transmitting under extreme communication constraints. This section assesses four key signal processing architectures: (1) voltage beamforming (BF), (2) direct imaging using an FFT-based beamformer called E-field Parallel Imaging "correlator" (EPIC), (3) traditional cross-correlation (XCorr), and (4) FFT-based correlator (FFTCorr)—with a focus on their relative data rates and on-site computational demands across multiple array configurations. Our analysis shows that **XCorr architectures consistently incur data rates 20–40 times higher** than BF, EPIC, or FFTCorr approaches. In terms of computational load, the advantages of EPIC become even more prominent. For both the full FarView layout and the compact core configuration, **EPIC requires significantly less processing power** than correlator-based methods. Overall, the analyses strongly indicate that **EPIC provides the most balanced solution**, minimizing both data transmission requirements and on-site compute resources from an operational perspective.*

### 8.1 Array Layout

We consider two nominal array configurations and examine each using their core and full layouts, which we denote as *FarView*-4×4 and *FarView*-3×3, and *FarView*-4×4-core and *FarView*-3×3-core. The core layouts are described in detail in Section 3.3 and in [59]. The outrigger specifications for both these layouts are loosely modeled based on the original description of *FarView* [58].

We parametrize the array layout using a number of array elements each of size, $D_e$, with $N_{eps}$ elements grouped into a station of size $D_S$, and $N_S$ stations are further arranged in an array of size, $D_A$. Table 1 lists the nominal values of the array parameters for the layouts of *FarView* considered here. In the case of *FarView*-4×4 and *FarView*-3×3 and their cores, the subarray consisting of 4×4 and 3×3 dipoles, respectively, forming the unit elements. The operating wavelength is denoted by $\lambda$, which is nominally assumed to be 10-m (30 MHz).

### 8.2 Data Processing Architectures

The architectures that we considered have been discussed in detail in Thyagarajan [79]. Here, we present a summary of the essential details.

#### 8.2.1 Images using Voltage Beamforming (BF)

In this architecture, the intra-station beamformed voltages are coherently beamformed together at the inter-station stage towards specific directions using a two-dimensional discrete Fourier transform. The angular resolution of the inter-station beamforming will correspond to the dimensions of the station layout within the array as $\sim (\lambda/D_A)^2$, and we consider all station beams that can be formed over the full field of view, $\sim (\lambda/D_e)^2$. This architecture requires near real-time calibration.

TABLE 1: SUMMARY OF FARVIEW ARRAY LAYOUT PARAMETERS.

| Array name | $D_A/\lambda$ [a] | $N_s$ | $D_s/\lambda$ [a] | $N_{eps}$ | $D_e/\lambda$ [a] | $f_A$ [b] | $f_s$ [c] |
|---|---|---|---|---|---|---|---|
| FarView-4x4 [d,e] | 1400 | $13^2 \times 6^2$ | 5 | $4^2$ | 1 | 0.078 | 0.64 |
| FarView-3x3 [d,e] | 1400 | $13^2 \times 8^2$ | 3.75 | $3^2$ | 1 | 0.078 | 0.64 |
| FarView-4x4-core [e] | 360 | $12^2 \times 6^2$ | 5 | $4^2$ | 1 | 1 | 0.64 |
| FarView-3x3-core [e] | 360 | $12^2 \times 8^2$ | 3.75 | $3^2$ | 1 | 1 | 0.64 |

a Nominal operating wavelength of the array is $\lambda = 10$ m. Results may differ at other wavelengths depending on aperture efficiency, filling factor, etc.
b Array filling factor, $f_A = N_s (D_s/D_A)^2$.
c Station filling factor, $f_s = N_{eps} (D_e/D_s)^2$.
d Refer to Smith & Pober [59].
e Refer to Polidan et al. [58] for layout of outliers.

### 8.2.2 Images Using E-field Parallel Imaging 'Correlator' (EPIC)

E-field Parallel Imaging 'Correlator' (EPIC; [80]) is an FFT-beamforming framework applied using the measured voltages but generalized to arbitrary array layouts using a gridding kernel [80]. It does not use correlations between antennas unlike a conventional correlator architecture. EPIC can be applied to one or both of intra- and inter-station processing. The output of EPIC consists of images filling the entire field of view at the native angular resolution of the array. The use of FFT and the avoidance of a correlator can be advantageous for large aperture arrays that are densely packed. This architecture also requires near real-time calibration, for which schemes are available [81].

### 8.2.3 Visibilities from Correlator (XCorr)

This is a traditional radio interferometry architecture in which the voltages from pairs of antennas are correlated to produce visibilities. The visibilities can be used in several ways in post-processing to produce the desired products. This architecture can be expensive if the array consists of many elements because the number of antenna pairs in the correlation process scales as $N_s^2$. When these visibilities are gridded and Fourier transformed to make images, it is denoted as XFFT.

### 8.2.4 Visibilities from Gridding (FFTCorr)

An alternative to the visibilities produced by the correlator architecture is to produce gridded visibilities. The gridding can potentially reduce the data volume if the *uv*-coverage has significant redundancy. Gridding of visibilities can be achieved either by gridding the measured visibilities from the correlator or using inverse Fourier transform of images created from the EPIC or beamforming architectures. The former allows more freedom of weighting schemes that can be applied in post-processing whereas the former has more computational efficiency afforded using an FFT. Whether it is explicit gridding of the correlator's visibilities or through inverse FFT of the images, the original visibilities that were used in either of these architectures will necessarily have to be calibrated.

The *FarView*-4×4 and *FarView*-4×4-core layouts have very similar requirements from data rate and computational cost perspectives relative to their *FarView*-3×3 counterparts. Thus, we only present the analysis for *FarView*-4×4 and *FarView*-4×4-core layouts hereafter.

## 8.3 Sample Rates

Table 2 lists the dependence of the outputs from various data processing architectures on the array layout parameters and the accumulation interval.

Figures 24 and 25 show the sample rates from the different processing architectures for the layouts considered as their nominal parameters are varied. Each panel denotes the variation of the sample rate with the parameter specified on the x-axis while the other parameters are kept fixed at their nominal values indicated by the vertical cyan lines.

All layouts exhibit some common trends. For example, the pixel-based architectures, namely, BF, EPIC, and FFTCorr are relatively insensitive to $N_s$, $N_{eps}$, and $D_s$ because the number of pixels in the image or the Fourier grid corresponding to filling the field of view, $\sim(\lambda/D_e)^2$, with independent resolution elements, $\sim(\lambda/D_A)^2$, are independent of these parameters. The number of pixels or the size of the Fourier grid scale as $\sim(D_A/D_e)^2$. The data rate of XCorr outputs scales as $\sim N_s^2$ accounting for the visibilities from the correlator. The correlator needs to produce visiblities phased to every station beam covering the field of view, and thus scales as $\sim(D_s/D_e)^2$. All data processing architectures depend inversely on $t_{acc}$.

**TABLE 2: SAMPLE RATE BUDGET OF INTER-STATION COHERENT DATA OUTPUT**

| Architecture | Data Rate (bytes per accumulation)[a] |
|---|---|
| Voltage Beamforming (BF) | $N_r N_v n_p^2 (D_A/D_e)^2 / t_{acc}$ |
| Direct Imaging (EPIC) | $N_r N_v n_p^2 (\gamma_A D_A/D_e)^2 / t_{acc}$ [b] |
| Correlator Visibilities (XCorr) | $2 N_r n_p^2 (D_s/D_e)^2 [N_s(N_s-1)/2] / t_{acc}$ |
| Gridded Visibilities (FFTCorr) | $2 N_r N_v n_p^2 [((\gamma_A D_A/D_e)^2 + 1)/2] / t_{acc}$ [bc] |

a $N_r$, $N_v$, and $n_p$ denote the number of bytes used to represent a real number, the number of frequency channels, and number of polarization states measured, respectively.
b $\gamma_A$ denotes padding factor for array-level FFT used for EPIC and gridded visibilities. Here, $\gamma_A = 2$ along each dimension to achieve identical image and pixel sizes as that from XFFT.
c The value of unity inside the parenthesis applies only if autocorrelations are measured from the gridded visibilities, which consists of $n_p^2$ real numbers.

Although there are apparent differences at the level of a factor of few between pixel-based architectures such as BF, EPIC, and FFTCorr, these are superficial, arising primarily due to the use of a padding factor in the FFT-based architectures like EPIC and FFTCorr. However, these differences can be equalized by smoothing the data to output the same pixels as from the BF architecture. Thus, these differences can be neglected, and the FFT-based architectures, EPIC, and FFTCorr, can be assumed to produce essentially the same data rates as the BF architecture.

In all the nominal array layout parameters considered, the XCorr architecture requires ~20 to 40 times higher data rate than the BF, EPIC, and FFTCorr architectures. Thus, pixel-based architecture has an advantage over the correlator architecture from the viewpoint of data rate.

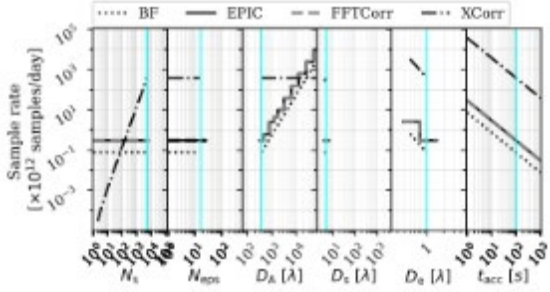
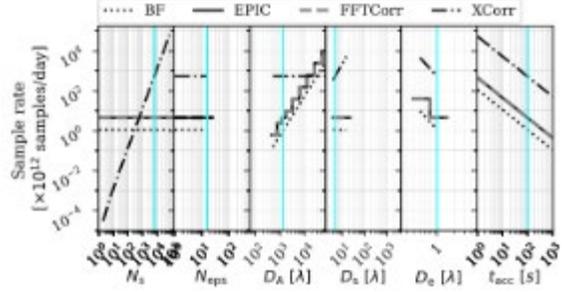

**Figure 24.** Sample rates for *FarView*-4x4.  **Figure 25.** Sample rates for *FarView*-4x4-core.

## 8.4 Computational Costs

The dependence of computational costs of various data processing architectures on the array layout parameters and the accumulation interval is listed in Table 2 of [79].

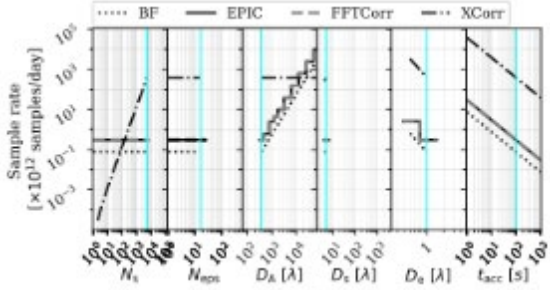
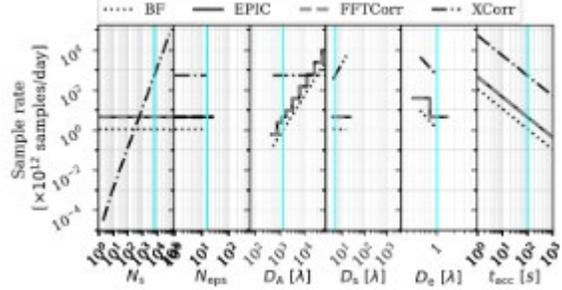

**Figure 26.** Compute costs for *FarView*-4x4.  **Figure 27.** Compute costs for *FarView*-4x4-core.

Figures 26 and 27 show the computational costs for the different processing architectures for the layouts considered as their nominal parameters are varied. Each panel denotes the variation of the sample rate with the parameter specified on the x-axis while the other parameters are kept fixed at their nominal values indicated by the vertical cyan lines.

The line styles denote the two-stage data processing architecture, while the gray and black lines correspond to the station-level processing being a voltage beamformer (BF) and EPIC, respectively. The black and gray lines are virtually indistinguishable in all cases indicating that the inter-station processing costs dominate the overall cost budget relative to the intra-station costs. Thus, we restrict the discussion below to comparing the inter-station computational costs.

The compute cost for BF scales linearly with $N_s$ while that for EPIC is relatively insensitive to $N_s$ because EPIC's cost primarily depends on the FFT, which in turn depends on the grid size rather than on $N_s$. The cost of correlator-based architectures like XBF and XFFT scale as $N_s^2$.

All the inter-station architectures are essentially insensitive to $N_{eps}$ and $D_{eps}$ as these parameters are relevant mainly to intra-station processing only. The BF architecture will scale as $\sim (D_A/D_s)^2$, and hence its cost per voxel is insensitive to $D_A$ and $D_s$. The FFT cost per voxel in EPIC has only a weak dependence on $D_A$ and $D_s$. The correlator-based architectures, XBF and XFFT, dominated by the correlations, will be independent of $D_A$ and $D_s$, and hence their per-voxel cost scale as $\sim (D_A/D_s)^{-2}$. In the case of XBF, at larger values of $D_A$, the discrete Fourier Transform (DFT) starts dominating the cost relative to the correlator, and hence the per-voxel cost flattens. For the timescale range considered, all architectures except XBF remain flat with $t_{acc}$. The XBF also flattens on large timescales but at smaller timescales, it scales as $t_{acc}^{-1}$ on account of being dominated by the DFT cost.

These trends hold for the nominal values of both the full and core *FarView* layouts considered. In both cases, the EPIC architecture for inter-station processing has a computational advantage over other architectures. At the nominal values for *FarView*-4×4, EPIC is advantageous by a factor of few relative to the correlator architectures. However, the differences are more pronounced for *FarView*-4×4-core where EPIC is ≥50 times more efficient than correlator architectures.

Additional work is needed to determine the data transmission rate to Earth given the EPIC architecture suggested here for *FarView*. Previously, we estimated data rates of 2.4-5.8 TB/day [58] using a generic subarray architecture which will be further refined in the next phase of development.

## 9.0 Conclusions, Acknowledgements, and Future Work

This report summarizes the results of multi-year design studies and science analyses for the *FarView* low-frequency radio array on the lunar far side. The project was supported by NASA Innovative Advanced Concepts (NIAC) Phase I and II awards (grant 80NSSC23K0965) and Lunar Resources, Inc.

The primary science driver for *FarView* is to enable pioneering observations of the unexplored Cosmic Dark Ages and early Cosmic Dawn (redshift $z \approx$ 30–100) - a period identified by the **Astro2020 Decadal Survey** as the "Discovery Area for Cosmology." With its unprecedented spatial and spectral resolution, *FarView* will probe the physics of cosmic inflation, map large-scale structure in its linear regime, and investigate the nature of dark matter beyond the standard cold dark matter model. These observations rely on detecting the redshifted 21-cm hyperfine transition of neutral hydrogen at frequencies below 50 MHz, achievable only from the Moon's far side—a uniquely radio-quiet environment free from Earth's ionosphere and anthropogenic radio frequency interference.

Meeting these requirements demands a large-scale interferometric array. Numerical simulations show that a fiducial design with ~100,000 dipole antennas - 80,000 densely packed in a 4-km core and 20,000 in a 14-km halo - can achieve a 10σ detection of the Dark Ages power spectrum over five years (50% duty cycle), presuming foregrounds can be measured accurately. The hierarchical architecture of phased subarrays and subarray clusters balances sensitivity and feasibility. Unlike ground-based Epoch of Reionization experiments, *FarView* cannot rely on foreground avoidance; instead, it will employ high-resolution modeling and subtraction, leveraging the lunar environment's stability.

Beyond cosmology, *FarView* enables transformative science across multiple domains:

- **Heliophysics and Space Weather:** Imaging Type II and III solar radio bursts to constrain particle acceleration and turbulence in the inner heliosphere, advancing NASA's *Moon-to-Mars objectives* and recommendations from the **Solar & Space Physics Decadal Survey**.

- **Exoplanetary Science:** Detecting coherent radio bursts from other stellar systems and magnetized exoplanets, providing direct measurements of magnetic fields critical for habitability. *FarView* will provide new constraints on space weather in nearby exoplanet systems. Interactions of stellar winds, flares, CMEs, and other forms of stellar mass loss are vital to constraining exoplanetary atmospheres and the ISM as noted by the *Astro2020 Decadal* survey.

- **Galactic Astrophysics:** Mapping bremsstrahlung absorption to produce the first 3D tomographic survey of cosmic ray distribution in the Milky Way. *FarView* can fill a key knowledge gap in our understanding of feedback and structure formation in the Milky Way as identified in the *Astro2020 Decadal Survey*.

Operational challenges include managing extreme data volumes and transmission constraints. A trade study of signal-processing architectures identifies the FFT-based **EPIC** beamformer as the most efficient solution, minimizing computational load and data rates. Additional research is also underway on in-situ manufacturing using molten regolith electrolysis to fabricate antennas from lunar materials.

Finally, time is critical. The lunar far side's pristine radio environment is threatened by unintended electromagnetic emissions from lunar infrastructure and lunar-orbiting satellites. Preserving this resource and deploying prototypes of *FarView* promptly will secure a unique opportunity for breakthrough science that cannot be achieved elsewhere in the foreseeable future.